%% file: ASUP_V18.tex
\documentclass[journal]{IEEEtran}

\usepackage{silence}
\input{preamble.tex}
\newacronym{CRLB}{CRLB}{Cram\'er-Rao lower bound}
\newacronym{FIM}{FIM}{Fisher Information Matrix}
\newacronym{SVD}{SVD}{singular value decomposition}

\crefname{condition}{condition}{conditions}
\crefformat{condition}{condition~(#2#1#3)}
\crefrangeformat{condition}{conditions~(#3#1#4) to~(#5#2#6)}
\crefmultiformat{condition}{conditions~(#2#1#3)}{ and~(#2#1#3)}{, (#2#1#3)}{ and~(#2#1#3)}

\newcommand{\inv}[1]{\left({#1}\right)^{-1}}

\newcommand{\wtil}[1]{\widetilde{#1}}
\newcommand{\tbP}{\widetilde{\mathbf P}}
\newcommand{\tbJ}{\widetilde{\mathbf J}}

\newcommand{\subs}[2]{\left[{#1}\right]_{#2}}
\newcommand{\tb}[1]{\widetilde{\mathbf{#1}}}
\newcommand{\bslash}[1]{\backslash {#1}}
\newcommand{\Null}[1]{\mathrm{Null}\left({#1}\right)}

\begin{document}

\title{Arbitrarily Strong Utility-Privacy Tradeoff in Multi-Agent Systems}
\author{Chong~Xiao~Wang, Yang~Song, Wee~Peng~Tay,~\IEEEmembership{Senior~Member,~IEEE}%
\thanks{
This work was supported in part by the Singapore Ministry of Education Academic Research Fund Tier 2 grant MOE2018-T2-2-019 and by A*STAR under its RIE2020 Advanced Manufacturing and Engineering (AME) Industry Alignment Fund – Pre Positioning (IAF-PP) (Grant No. A19D6a0053). The computational work for this article was partially performed on resources of the National Supercomputing Centre, Singapore (https://www.nscc.sg). A preliminary version of this paper was presented at the Proc. IEEE Global Conf. on Signal and Information Processing, Anaheim, USA, Nov, 2018. The authors are with the School of Electrical and Electronic Engineering, Nanyang Technological University, Singapore. E-mails:wangcx@ntu.edu.sg, songy@ntu.edu.sg, wptay@ntu.edu.sg.
}%
}

%\markboth{IEEE TRANSACTIONS ON INFORMATION THEORY,~Vol.~, No.~, ~2012}%
%{Tay: xxxx}

% The paper headers
%\markboth
%   {To be submitted... }
%   {Tay \MakeLowercase{\textit{et al.}}: }

% make the title area
% Don't write page number 0 to the cover page.
\maketitle \thispagestyle{empty}

% Put abstract and the paper's body in a new page, page 1.
%\newpage
%\setcounter{page}{1}

%---------------------------------------------------------------------------%
%                           Abstract and key words                          %
%---------------------------------------------------------------------------%
\begin{abstract}
Each agent in a network makes a local observation that is linearly related to a set of public and private parameters. The agents send their observations to a fusion center to allow it to estimate the public parameters. To prevent leakage of the private parameters, each agent first sanitizes its local observation using a local privacy mechanism before transmitting it to the fusion center. We investigate the utility-privacy tradeoff in terms of the Cram\'er-Rao lower bounds for estimating the public and private parameters. We study the class of privacy mechanisms given by linear compression and noise perturbation, and derive necessary and sufficient conditions for achieving arbitrarily strong utility-privacy tradeoff in a multi-agent system for both the cases where prior information is available and unavailable, respectively. We also provide a method to find the maximum estimation privacy achievable without compromising the utility and propose an alternating algorithm to optimize the utility-privacy tradeoff in the case where arbitrarily strong utility-privacy tradeoff is not achievable.
\end{abstract}


%---------------------------------------------------------------------------%
%                              Introcution                                  %
%---------------------------------------------------------------------------%
\section{Introduction}
The increasing number of multifarious sensing and monitoring applications installed in mobile phones, offices and public facilities has led to the proliferation of various services based on network data analytics \cite{AKySuSan:J02, CaaKemPed:J17, TanJiTay:J18, YanZhoTay:J18}. Due to the limited data that a single sensor or agent can observe because of its geographical placement or location, agent type and computation capabilities, multi-agent networks \cite{ChaVee:03,TayTsiWin:J07,TayTsiWin:J08b,KreTsiZou:09,TayTsiWin:J09a,Tay:J12,ZhaChoPez:13,Tay:J15,HoTayQue:J15} are often deployed to overcome the insufficiency of data retrieved from a single agent. Agent fusion or sense-making is then used to combine the sensory data derived from disparate sources to reduce the inference uncertainty. However, aggregation of data poses a higher risk of privacy leakage. For example, social network users in a community may share their personal opinions or experiences about different products over time. While the aggregated data can be useful feedback for a company to improve its own product, with a database recording the preferences of each user for multiple products over a period of time, one can infer personal traits and other sensitive attributes like the gender and income level of a user. Furthermore, studies \cite{IraWebLi:J2011} have shown that information from a user's friends on a social network can accurately reveal the user's marital status, location, sexual orientation or political affiliation. Therefore, it is imperative that data from each source is sanitized to reduce privacy leakage before revealing it to the public. 

In this paper, we consider a multi-agent network where each source, node or agent (for convenience, we call this an agent throughout the paper) makes a noisy observation, which is linearly related to a set of system parameters. A set of public and private parameters are defined to be linear maps of the system parameters. We assume the agents send their observations to a fusion center to allow it to infer the public parameter with high fidelity \cite{BehEltJaf:J2014,XiaoCuiLuo:J2008,SpeFisJoh:J2008,San:C2013}. The agents however want to keep the private parameters secret. To achieve this, each agent sanitizes its observation before sending to the fusion center. In this paper, we consider two sanitization methods: linear compression and noise perturbation. Since a trusted third-party who can help perturb the agent observations in a centralized manner does not exist or is impractical in many applications, a decentralized\footnote{The term ``decentralized'' refers to the data sanitization process at each agent, which is independent of the other agents.} sanitization scheme is considered where each agent performs its local sanitization independently.

\subsection{Related Work}
We can classify privacy into two types: data privacy and inference privacy \cite{CalFaw:C12, SunTayHe:J18, SunTay:C17, SunTay:J20b}. Data privacy often refers to protecting access to the raw data while inference privacy refers to the prevention of illegitimately inferring sensitive information \cite{CalFaw:C12, SunTayHe:J18}. Homomorphic encryption is a classical method for data privacy \cite{PlaSusZha:J13}. However, such an approach is unable to hide the sensitive information contained in the encrypted data. Differential privacy\cite{NyPap:J14,YeBar:J2018} ensures the indistinguishability of the query records in a database. However, differential privacy only deals with a source alphabet with finite support, and thus does not apply to the privacy in estimation theory. The other privacy metrics that have been extensively used in inference privacy include mutual information (entropy) privacy, average information leakage and maximum information leakage \cite{WanYinZha:J16}. The privacy we consider in this paper belongs to the category of inference privacy.

Privacy-preserving estimation and detection from a single data source or agent has been well-studied by using information-theoretic approaches. The paper \cite{SanRaj:J2013} presented an information-theoretic framework that ensures the utility of the data source while providing necessary privacy guarantees in a database associated with a statistical model. The following works are based on a general privacy statistical framework: Two random variables $(X,Y)$ are assumed to be associated with a given joint distribution, and a user observes $Y$ and wants to disclose to another user as much information about $Y$ as possible while limiting the amount of information revealed about $X$ \cite{AsoDiaLin:J15}. To achieve that, the data is transformed before being disclosed, according to a probabilistic privacy mapping. The paper \cite{CalFaw:C12} introduced two privacy metrics, namely average information leakage and maximum information leakage, and showed that optimal privacy-accuracy tradeoff can be cast as modified rate-distortion problems. In \cite{AsoDiaLin:J17,AsoAlaLin:C16}, the authors formulated the privacy-utility tradeoff in terms of the smallest normalized minimum mean-squared error. Furthermore, the references \cite{WanCal:C2017,WanLis:J2018} characterized the fundamental performance limits of privacy-assuring mechanisms from an estimation theoretic perspective, and developed data-driven privacy mechanisms that provide estimation-theoretic guarantees \cite{CalMak:J2017}. The papers \cite{HsuAsoSal:J2018,MakSalFaw:C2014} introduced a log-loss metric to measure privacy and utility and linked the privacy funnel method to the information bottleneck method \cite{NatPerWil:J2000}. In \cite{LiuKhiMah:C17, LiuLeeKhi:J17, LiKhiMah:J18, GiacGunPoo:C15}, the authors investigated the utility-privacy tradeoff quantified by mutual information in smart metering using a rate-distortion approach. All the above-mentioned works only deal with single entry data and do not generalize immediately to the multi-agent setting where decentralized sanitization mechanism is required. One of the underlying presumptions of these works is that a single user owns the data, while this paper considers the case where data is distributed among multiple users.

Several papers have addressed the issue of privacy protection under the hypothesis testing framework in a multi-agent system from different perspectives. The papers \cite{LiOce:J2015,LiOec:J2017} investigated the privacy leakage problem in an eavesdropped distributed hypothesis test network from the Bayesian detection perspective. Under a similar decentralized detection framework, the papers \cite{SunTayHe:J18, SunTay:C16} proposed a nonparametric learning approach to design local privacy mappings to distort each agent's observation, thus preventing the fusion center from using its received information to accurately infer the private hypothesis. The paper \cite{SunTay:J20a} considered ways to achieve robust information privacy for a set of private hypotheses while \cite{HeTayHua:J19} proposed a multi-layer agent network where non-linear fusion is applied. In contrast to these papers, which focus on the protection of a private hypothesis in a \emph{hypothesis testing} framework, we consider in this paper the privacy protection of a set of parameters in a parameter \emph{estimation} framework. An example is in the deployment of various body sensors (agents) for evaluating a user's health condition \cite{AleErs:J10}. With the raw observations sent from the sensors, a service provider can not only analyze the user's health condition but can also infer some sensitive information about the user such as her location and personal preferences. Therefore, preserving the privacy of certain sensitive parameters associated with the sensor data is an important requirement.

The references \cite{LiuKhiMah:C17, AsoAlaLin:C16} proposed to add random noise to perturb raw measurements, while \cite{EmaMil:C13,Diamantaras2016,KunSPM2017,KunJFI2017,Al2017,ChaChaKun:C17,SonWanTay:C18,SonWanTay:J20} investigated the use of compressive linear mappings that transform the raw measurements to a lower dimensional space. Preserving the privacy of individual entries of a database with constrained additive noise was considered in \cite{FarSan:C17} where a measure of privacy using the Fisher information matrix \cite{LiuCheFar:J16} was developed. These works did not explore the connection between adding random noise and compressive linear transformations as privacy mechanisms. In this paper, one of our contributions is to clarify the relationship between these two mechanisms.

\subsection{Our Contributions}
In this paper, we consider the case where agents in a network send sanitized observations to a fusion center to allow it to infer a set of public parameters, while preventing it from estimating a set of private parameters better than a predefined accuracy. Our main contributions are the following:
\begin{enumerate}
\item We make explicit the relationship between additive random noise and linear compression as privacy mechanisms under the \gls{CRLB} privacy framework. We show that the \gls{CRLB}s for estimating the private parameters under the linear compression mechanism form the boundary of the set of \gls{CRLB}s under noise perturbation.
\item We introduce the notion of arbitrarily strong utility-privacy tradeoff (ASUP), and derived necessary and sufficient conditions under which this is achievable. In addition, we propose a method to find the maximum privacy that can be attained while maintaining perfect utility under a constraint on the noise perturbation power.
\item In the case where ASUP is not achievable, we propose an alternating optimization algorithm to find a sanitization to achieve an optimal utility-privacy tradeoff.
\end{enumerate}

A preliminary version of this work was presented in \cite{WanSon:C2018} in which the noise perturbation method was used to protect the private parameters, while allowing the inference of the public parameters. The present paper delves into the analysis of the privacy and utility tradeoff in multi-agent systems with decentralized sanitization schemes and clarifies the relationship between noise perturbation and linear compression. New theoretical insights and methods are also presented.

The rest of this paper is organized as follows. In \cref{sect:problem_formulation}, we present our problem formulation and assumptions. In \cref{sec:pp_transformation}, we investigate the relationship between additive random noise and linear compression under our CRLB privacy framework. In \cref{sect:asup}, we present necessary and sufficient conditions for ASUP. In \cref{sect:max_privacy}, we consider the case where ASUP is not achievable and investigate maximum privacy under perfect utility with power constraint. An alternating optimization algorithm to optimize privacy under perfect utility is presented in \cref{sect:sequential_optimization}. We present numerical simulation results in \cref{sect:simu_discussion} and conclude in \cref{sect:conclusion}.

\emph{Notations:}
We use $\bbR$ to denote the set of real numbers, and $\bbS^n, \bbS_{+}^n$ to denote the set of $n \times n$ positive semi-definite matrices and positive definite matrices, respectively. We use $\T$ to represent matrix transpose. The notation $\bzero_N$ is the $N\times N$ zero matrix, and $\bI_N$ is an $N\times N$ identity matrix. We write $\bA \succeq \bB$ if $\bA - \bB$ is positive semi-definite, while $\bA \succ \bzero$ means $\bA$ is positive definite.  We use $\diag(\cdot)$ to denote the block diagonal operation and $\trace{\cdot}$ the trace operation. We use $[\bP]_{\calP,\calQ}$ to denote the sub-matrix of the matrix $\bP$ consisting of the entries $\bP(i,j)$ for all $(i,j)\in \calP\times\calQ$ and use $[\bP]_{:,\calQ}$ and $[\bP]_{\calQ,:}$ to denote the sub-matrix of the matrix $\bP$ consisting of, respectively, the columns and rows, indexed by $\calQ$. The rank of the matrix $\bA$ is $\rank(\bA)$ and $\Null{\bA}$ denotes its null space. The Gaussian distribution with mean $\mu$ and variance $\sigma^2$ is denoted as $\N{\mu}{\sigma^2}$, and the uniform distribution on the interval $\left[a,b\right]$ is denoted as $\Unif{a,b}$.

%---------------------------------------------------------------------------%
%                          Problem Formulation                              %
%---------------------------------------------------------------------------%
\section{Problem formulation} \label{sect:problem_formulation}
In this section, we present our system model and assumptions. Consider a multi-agent network consisting of the agents $i = 1, \ldots, S$ and a fusion center. Each agent $i$ makes a noisy observation $\by_i\in {\bbR}^{N_i}$ about a system parameter $\bx \in \bbR^{L}$. The observation model for agent $i$ is given by
\begin{align*}
	{\by}_i = {\bH}_i {\bx} + {\bn}_i,
\end{align*}
where ${\bH}_i \in {\bbR}^{N_i \times L}$ is the observation model matrix, and ${\bn}_i \sim \N{\bzero_{N_i}}{\bR_i}$ is the measurement noise. Here, $N_i$ may be smaller than $L$, hence each agent $i$ may not be able to infer the system parameter $\bx$ based on its own observation $\by_i$. Furthermore, even if an agent $i$ is able to estimate $\bx$ from its local observation $\by_i$, the fusion center with access to all agents' observations achieves a higher accuracy than each individual agent. 

In this paper, we consider the case where the fusion center aims to estimate $\bu = \bU \bx \in \bbR^{U}$, where $\bU \in \bbR^{U \times L}$ and $U>0$, based on information it receives from all agents in the network. However, the agents also want to prevent the fusion center from inferring a set of private parameters $\bg_i = \bG_i \bx \in \bbR^{G_i}$, where $\bG_i \in \bbR^{G_i \times L}$, $i = 1,\ldots,S$. We assume that at least one $\bG_i$, $i=1,\ldots,S$, is non-zero. Otherwise our problem formulation reduces to the case without privacy consideration. 

Stacking up all agents' measurements, we have
\begin{align}
	\by = {\bH \bx} + \bn,
\end{align}
where $\by = \left[ {\by}_1\T, \ldots , {\by}_{S}\T \right]\T \in {\bbR}^N$, 
${\bH} = \left[ {\bH}_1\T, \ldots , {\bH}_{S}\T \right]\T \in {\bbR}^{N \times L}$, 
${\bn} = \left[ {\bn}_1\T, \ldots , {\bn}_{S}\T \right]\T \in {\bbR}^N$,
and $N = \sum_{i=1}^{{S}} N_i$. Let $\calS_i=\{\sum_{k=1}^{i-1}N_k+1,\ldots,\sum_{k=1}^{i}N_k\}$ be the index set corresponding to agent $i$. Note that the measurement noise ${\bn}_i$ and ${\bn}_j$ for any two different agents $i$ and $j$ are not necessarily independent of each other. We assume that $\bn \sim \calN \left(\bzero_N, \bR\right)$, where the block diagonal part of $\bR$ is equal to $\diag \left( {\bR}_1, \ldots, {\bR}_{S} \right)$ and the off-diagonal entries are the noise correlations between agents. 

\begin{Example}
Consider an audio system consisting of $S$ microphone arrays placed at different spatial locations. Let $\subs{\bx}{k}$ represent the narrow band signal of the $k$-th person's speech in a group, and $\by_i=\bH_i\bx+\bn_i$ be the mixed signals received by the $i$-th microphone array, where $\bn_i$ is a complex Gaussian noise, and $\bH_i$ is a mixing matrix, where $\subs{\bH_i}{j,k}=\alpha_i\mathrm{e}^{\iu\theta_{ijk}}$ with $\iu=\sqrt{-1}$, $\alpha_i$ being the gain and $\theta_{ijk}$ being the phase shift of the received signal of the $k$-th person's speech at the $j$-th sensor of the $i$-th microphone array. The microphone arrays send the received signal to a cloud service to decode the speech of a subgroup of persons in the index set $\calP$. Meanwhile, we wish to protect the speech signals of people in another distinct group indexed by $\calQ$ from the cloud service. Accordingly, $\bU$ is diagonal matrix with 1's on the diagonal at the row indices in the index set $\calP$ and zero everywhere else, and for each $i=1,\ldots,S$, $\bG_i$ is a diagonal matrix with 1's on the diagonal at the row indices in the index set $\calQ$ and zero everywhere else.
\end{Example}

%<*tag:knownpara>
We assume all system parameters and model (including $\bH$, $\bR$, $\bU$, $\bG$, etc.) are known to the fusion center.
%</tag:knownpara>
To protect the private parameters $\{\bg_i: i=1,\ldots,S\}$ from being inferred by the fusion center based on the collective measurements, each agent sanitizes its local observation before transmitting to the fusion center. Let $\calT(\by)=[\calT_1(\by_1)\T,\ldots,\calT_{S}(\by_{S})\T]\T$ denote the sanitized information received at the fusion center, where $\calT$ belongs to a predefined class of sanitization mechanisms. We measure the utility and privacy by the \gls{CRLB}s for estimating the public parameter $\bu$ and private parameters $\{\bg_i : i=1,\ldots,S\}$, respectively. 
%<*tag:CRLB>
Recall that the \gls{CRLB} \cite{Cra:B1945,BicDok:B77} is a lower bound on the variance of an unbiased estimator. Suppose $p_{\by|\bx}$ is a probability density function of a random variable $\by$ conditioned on $\bx$, and $\bx\sim p_{\bx}$ (prior information). For any unbiased estimator $\hat{\bx}(\by)$ of $\bx$ based on $\by$, we have $\E[\left(\hat{\bx}(\by)-\bx\right)\left(\hat{\bx}(\by)-\bx\right)\T]\succeq\left(\bJ_{\bx}+\bJ_0\right)^{-1}$, where
\begin{align*}
\bJ_{\bx}
=-\E[\frac{\partial^2\log(p_{\by|\bx}(\by|\bx))}{\partial{\bx}^2}],\
\bJ_0
=-\E[\frac{\partial^2\log(p(\bx))}{\partial{\bx}^2}]
\end{align*}
with $\left(\bJ_{\bx}+\bJ_0\right)^{-1}$ known as the \gls{CRLB} for estimating $\bx$. If $\bx$ is a deterministic parameter ($\bJ_0=\bzero$), we have $\cov(\hat{\bx}(\by))\succeq\bJ_{\bx}^{-1}$ for any unbiased estimator $\hat{\bx}(\by)$. In other words, no unbiased estimator can outperform the \gls{CRLB} in terms of error variance.
%</tag:CRLB>

Denoting the \gls{CRLB} of the system variable $\bx$ before sanitization as $\bP_\bx$ and after sanitization as $\wtil{\bP}_{\bx}$, the \gls{CRLB}s for the public parameter $\bu$ before and after sanitization at the agents are
\begin{align*}
	\bP_{\bu} &= \bU \bP_\bx \bU\T, \\
	\wtil{\bP}_{\bu} &= \bU \wtil{\bP}_{\bx} \bU\T,
\end{align*}
respectively. Similarly, the \gls{CRLB}s for the private parameter $\bg_i$ before and after sanitization at the agents are
\begin{align*}
	\bP_{\bg_i} &= \bG_i \bP_\bx \bG_i\T, \\
	\wtil{\bP}_{\bg_i} &= \bG_i \wtil{\bP}_{\bx}\bG_i\T,
\end{align*}
respectively.

The quantity $\trace{\bP_\bu}$ gives a lower bound for the sum error variance for estimating every component of $\bu$. For each sanitization function $\calT$, we define the system utility function for the public parameter $\bu$ as 
\begin{align} \label{eq:Utility_Loss}
u(\calT) = 1 - \frac{\trace{\wtil{\bP}_{\bu}}}{\trace{\bP_\bu}} \leq 0,
\end{align}
which is the negative of the percentage increase in sum error variance lower bound for estimating $\bu$ due to privacy sanitization. We define the system privacy function for the private parameter $\bg_i$, $i=1,\ldots,S$, as
\begin{align} \label{eq:Privacy_Gain}
p_i(\calT) = \frac{\trace{\wtil{\bP}_{\bg_i}}}{\trace{\bP_{\bg_i}}} - 1,
\end{align}
which is the percentage increase in sum error variance lower bound for estimating $\bg_i$ due to privacy sanitization. Since $\calT$ can be viewed as part of the estimation procedure, we have $\wtil{\bP}_\bx \succ \bP_\bx$. Therefore, $u(\calT) \leq 0$ and $p_i(\calT)\geq 0$ for all $\calT$. The utility and privacy functions as defined are thus intuitive: perturbation on the measurement decreases the utility and increases the privacy. Our goal is to
\begin{align}\tag{P0} \label{opt:P0}
\begin{aligned}
	\max_{\calT} \ & {u(\calT)}, \\
	{\rm s.t.}\ & p_i(\calT) \geq \epsilon_i,\ i = 1, \ldots, S, 
\end{aligned}
\end{align}
where $\epsilon_i \geq0$ is called the \emph{privacy threshold} for agent $i$. We summarize the notations introduced so far in \cref{tab:symbols} for the reader's convenience.
\begin{table*}[!t]
\centering
\caption{Summary of commonly-used symbols}
\label{tab:symbols}
\begin{tabular}{|C{0.1\textwidth}|L{0.6\textwidth}|}
	\hline
	Notation & Definition \\
	\hline \hline
	$\by_i$ &  The agent $i$'s measurements: $\by_i=\bH_i\bx+\bn_i\in\bbR^{N_i}$, where $\bx\in\bbR^L$ and $\bn_i\sim\N{\bzero_{N_i}}{\bR_i}$. \\
	\hline
	$\by$, $\calT(\by)$ & The raw measurements and sanitized measurements from all agents $1,\ldots,S$. \\
	\hline
	$\bR,\bH$ & The agents' measurements noise covariance matrix $\bR\in\bbS^N$ and observation matrix $\bH\in\bbR^{N\times L}$. \\
	\hline
	$\calS_i$ & The index set corresponding to agent $i$, i.e., $\subs{\bR}{\calS_i,\calS_i} = \bR_i$ and $\subs{\bH}{\calS_i,:}=\bH_i$. \\
	\hline
	$\bu,\bg_i,\bU,\bG_i$ & The public parameter $\bu=\bU\bx\in\bbR^U$ and the $i$-th private parameter $\bg_i=\bG_i\bx\in\bbR^{G_i}$. \\
	\hline
	$\bP_\bx$ & The \gls{CRLB} for estimating $\bx$ from the raw measurements $\by$. \\
	\hline
	$\tbP_{\bx},\tbP_{\bu},\tbP_{\bg_i}$ & The \gls{CRLB} for estimating $\bx$, $\bu$ and $\bg_i$, respectively, from sanitized data $\calT(\by)$. \\ 
	\hline
	$u(\calT),p_i(\calT)$ & The utility function for $\bu$ and privacy function for $\bg_i$, respectively, using sanitization function $\calT$. \\
	\hline
	$\epsilon_i$ & The privacy threshold assigned to agent $i$. \\
	\hline
	$\bPhi$ & If $\bJ_0=\bzero_L$, $\bPhi=\bR^{-1}-\bR^{-1}\bH\bP_\bx\bH\T\bR^{-1}$. Otherwise, $\bPhi=\inv{\bH\bP_0\bH\T+\bR}$. Cf.\ \cref{eq:decomposed_eq_no_prior_Phi,eq:decomposed_eq_with_prior_Phi}.  \\
	\hline
	$\bPsi$ & If $\bJ_0=\bzero_L$, $\bPsi=\bP_0\bH\T\bPhi$. Otherwise, $\bPsi=\bP_\bx\bH\T\bR^{-1}$. Cf.\ \cref{eq:decomposed_eq_no_prior_Psi,eq:decomposed_eq_with_prior_Psi}. \\
	\hline
	\end{tabular}
\end{table*}
%

%---------------------------------------------------------------------------%
%                            Privacy Mechanism                              %
%---------------------------------------------------------------------------%
\section{Decentralized privacy-preserving transformation}\label{sec:pp_transformation}
In this section, we study two sanitization schemes $\calT$ that are widely used in the privacy literature: 1) linear compression that reduces the dimension of the measurements \cite{KunSPM2017}, and 2) noise perturbation that adds random noise to the measurements \cite{LiKhiMah:J18}. We derive the relationship between these two schemes under CRLB-based privacy criteria like that in \cref{eq:Privacy_Gain}. We show that the noise perturbation scheme is equivalent to linear compression in an asymptotic sense.

To prevent the fusion center or potential adversarial agents from inferring the set of private parameters $\bg_1, \ldots, \bg_S$ through the measurement $\by$, we consider an affine transformation $\calT_i: \bbR^{N_i} \mapsto \bbR^{M_i}$ where $M_i \leq N_i$ to sanitize the measurement of agent $i$ to obtain:
\begin{align*}
	\calT_i(\by_i) = \bC_i \by_i + \bxi_i = \wtil{\bH}_i \bx_i + \wtil{\bn}_i + \bxi_i&,
\end{align*}
where 
$\bC_i \in \bbR^{M_i \times N_i}$ is called the compression matrix and $\bxi_i \in \bbR^{M_i}$ is a perturbation noise with $\bxi_i \sim \N{\bzero_{N_i}}{\bTheta_i}$. We also have $\wtil{\bH}_i = \bC_i \bH_i \in \bbR^{M_i \times L}$, $\wtil{\bn}_i = \bC_i \bn_i \in \bbR^{M_i}$ and $\wtil{\bn}_i \sim \N{\bzero_{N_i}}{\wtil{\bR}_i}$ with $\wtil{\bR}_i = \bC_i \bR_i \bC\T_i$. By collecting all the perturbed measurements, the measurement model at the fusion center can be summarized as follows:
\begin{align*}
	\calT(\by) = \bC \by + \bxi = \wtil{\bH} \bx + \wtil{\bn} + \bxi,
\end{align*}
where $\calT(\by) = \left[\calT_1(\by_1)\T, \ldots, \calT_S(\by_S)\T \right]\T \in \bbR^{M}$, $\tb{H} = \bC \bH$, $M=\sum_{i=1}^S M_i$, $\tb{n} \sim \N{\bzero_N}{\bC \bR \bC\T}$, $\bxi\T = \left[\bxi_1\T, \ldots, \bxi_{S}\T \right] \in \bbR^M$, $\bxi \sim \N{\bzero_N}{\bTheta}$, $\bC = \diag(\bC_1, \ldots, \bC_{S}), \bTheta = \diag(\bTheta_{1}, \ldots, \bTheta_{S})$.

Note that the global sanitization function $\calT$ is composed of independent local sanitization functions $\calT_1, \ldots, \calT_S$, the $i$-th of which applies a local linear compression matrix $\bC_i$ and additive noise $\xi_i$ with covariance $\bTheta_{i}$ to the $i$-th agent's measurement $\by_i$. Both the global compression matrix $\bC$ and additive noise covariance matrix $\bTheta$ have block diagonal form. There is no message exchange between agents during the sanitization process as this is prone to privacy attacks. 

From \cref{eq:Utility_Loss,eq:Privacy_Gain}, we can express the utility and privacy function \gls{wrt} $\bC$ and $\bTheta$ as
\begin{align} \label{eq:utility_function}
u(\bC, \bTheta) &= 1 - \frac{\trace{\bU \tbP_{\bx}(\bC, \bTheta) \bU\T}}{\trace{\bU \bP_\bx \bU\T}}, \\ \label{eq:privacy_function}
p_i(\bC, \bTheta) &= \frac{\trace{\bG_i \tbP_{\bx}(\bC, \bTheta) \bG_i\T}}{\trace{\bG_i \bP_\bx \bG_i\T}} - 1,
\end{align}
where
\begin{align} \label{eq:perturbed_CRLB}
\begin{split}
\tbP_{\bx}(\bC, \bTheta)
&= \tbJ_{\bx}(\bC, \bTheta)^{-1} \\
&= \inv{\bJ_0 + \bH\T \bC\T \inv{\bC\bR\bC\T + \bTheta} \bC\bH}.
\end{split}
\end{align}
Here, $\bJ_0$ is the \gls{FIM} of any prior information. We let $\bJ_0=\bzero_L$ if no prior information is available. There is no loss in generality if we restrict compression matrices $\bC_i$s to be square matrices, i.e., we consider $(\bC,\bTheta)$ to be chosen from the following sanitization parameter set
\begin{align}
\begin{aligned}
\calC = \Big\{ \left(\bC, \bTheta\right) \Big|\ & \bC = \diag(\bC_1, \ldots, \bC_{S}) \in \bbR^{N \times N}, \\
& \bTheta = \diag(\bTheta_{1}, \ldots, \bTheta_{S}), \\
& \bC_i \in \bbR^{N_i \times N_i}, \bTheta_{i} \in \bbS^{N_i},
i = 1,\ldots,S \Big\}.%
\end{aligned}
\end{align}
This is because the perturbed \gls{CRLB} remains unaltered by padding a non-square $\bC$ and $\bTheta$ with zeroes: $\tbJ_\bx(\bC,\bTheta) = \tbJ_\bx(\overline{\bC},\overline{\bTheta})$, where $\overline{\bC} = \begin{bmatrix}\bC \\ \bzero\end{bmatrix}$, $\overline{\bTheta}=\begin{bmatrix}\bTheta\ \bzero' \\ \bzero \end{bmatrix}$, and $\bzero$ and $\bzero'$ are $(N-M)\times N$ and $M\times (N-M)$ zero matrices respectively. 

In the following, we prove some properties of $\tbJ_\bx(\bC, \bTheta)$ with the domain $\calC$. We consider the metric spaces $\Real^{N\times N}\times\bbS^N$ and $\bbS^{L}$ endowed with the Frobenius norm so that $\calC$ is a subset of $\Real^{N\times N}\times\bbS^N$ and $\tbP_{\bx}(\cdot,\cdot)$ is a mapping from $\Real^{N\times N}\times\bbS^N$ to $\bbS^{L}$. Notions of metric properties like continuity are defined \gls{wrt} these metric spaces.

\begin{Lemma}\label{lem:tbJ_prop}
Let $(\bC,\bTheta)\in\calC$, where $\bC\in\Real^{N\times N}$ and $\bTheta\in\bbS^{N \times N}$. 
\begin{enumerate}[(i)]
	\item\label{it:continuous} $\tbJ_\bx(\bC,\bTheta)$ is a continuous function on $\calC$.
	\item\label{it:unitary} Suppose $\bA$ is either a unitary matrix or an invertible block diagonal matrix. Then, we have $\tbJ_\bx(\bA\bC,\bTheta)=\tbJ_\bx(\bC,\bA\T\bTheta\bA)$.
\end{enumerate}
\end{Lemma}
\begin{IEEEproof}
The first claim follows immediately from \cref{eq:perturbed_CRLB}. For the second claim, suppose first that $\bA$ is unitary. We then have
\begin{align*}
\tbJ_{\bx}(\bA\bC, \bTheta)
&= \bJ_0 + \bH\T \bC\T\bA\T \inv{\bA\bC\bR\bC\T\bA\T + \bTheta} \bA\bC\bH\\
&= \bJ_0 + \bH\T \bC\T\inv{\bC\bR\bC\T + \bA\T\bTheta\bA}\bC\bH \\
&= \tbJ_\bx(\bC,\bA\T\bTheta\bA).
\end{align*}
The proof for the case where $\bA$ is an invertible block diagonal matrix is similar and the proof is complete.
\end{IEEEproof}

\begin{Proposition}\label{prop:normalized}
\begin{enumerate}[(i)]
	\item\label{it:only_noise} Let $\calA = \set*{\tbJ_{\bx}(\bI_N, \bTheta)}{(\bI_N, \bTheta) \in \calC}$ and $\calB = \set*{\tbJ_{\bx}(\bC, \bTheta)}{(\bC, \bTheta) \in \calC}$. Then, $\calB \backslash \calA$ is the boundary of $\calA$.

	\item\label{it:normalized_noise} For any $(\bC, \bTheta) \in \calC$, there exists $(\bC', \bLambda_b) \in \calC$, where $\bLambda_b$ is a diagonal matrix with $\subs{\bLambda_b}{i,i} \in \{0, 1\}$, for all $i\geq 1$, such that 
\begin{align*}
\tbP_{\bx}(\bC', \bLambda_b) = \tbP_{\bx}(\bC, \bTheta).
\end{align*}
\end{enumerate}
\end{Proposition}
\begin{IEEEproof}
\begin{enumerate}[(i)]
	\item Since $\calA \subset \calB$, it suffices to show that for any $\bB \in\calB$, there exists a sequence in $\calA$ that converges to $\bB$. From \cref{lem:tbJ_prop}\ref{it:unitary}, we have $\tbJ_\bx(\bC,\bTheta) = \tbJ_\bx(\bI_N,\bC\T\bTheta\bC)$ with $\tbJ_\bx(\bI_N,\bC\T\bTheta\bC) \in \calA$ if $\bC$ is invertible. On the other hand, if $\bC$ is a singular matrix, we let the eigendecomposition of $\bC=\bQ\bLambda\bQ^{-1}$, where $\bQ$ is the square matrix whose columns are the eigenvector, and $\bLambda$ is the diagonal matrix whose diagonal elements are the corresponding eigenvalues. Denote $\bLambda = \diag(\bLambda', \bzero_A)$ with $\bLambda'$ containing all the non-zero eigenvalues and an appropriate $A$. From \cref{lem:tbJ_prop}\ref{it:unitary}, we have
	\begin{align} \label{overlinesigma}
	\tbJ_\bx(\bC,\bTheta)
	= \tbJ_\bx(\bQ\bLambda\bQ^{-1}, \bTheta)
	= \tbJ_\bx(\bLambda\bQ^{-1}, \bTheta'),
	\end{align}
	where $\bTheta' = \bQ\T\bTheta\bQ$. For any $\lambda>0$, let $\bLambda_\lambda = \diag(\bLambda', \lambda\bI_A)$, which is an invertible diagonal matrix. From \cref{lem:tbJ_prop}\ref{it:unitary}, we have
	\begin{align*}
	\tbJ_\bx(\bLambda_\lambda\bQ^{-1}, \bTheta')
	&= \tbJ_\bx(\bI_N, (\bLambda_\lambda\bQ^{-1})\T\bTheta'\bLambda_\lambda\bQ^{-1}) \in \calA,
	\end{align*}
	whose left-hand side converges to the right-hand side of \cref{overlinesigma} as $\lambda\to0$ since $\tbJ_\bx$ is continuous by \cref{lem:tbJ_prop}\ref{it:continuous}. The proof is now complete.

	\item Let the eigendecomposition of $\bTheta = \bQ\bLambda\bQ\T$. The diagonal matrix $\bLambda$ can be written as $\bLambda_{+}^{1/2}\bLambda_b \bLambda_{+}^{1/2}$, where $\bLambda_{+}$ and $\bLambda_b$ are diagonal matrices with $\subs{\bLambda_{+}}{i,i}=\subs{\bLambda}{i,i}$, $\subs{\bLambda_b}{i,i}=1$ when $\subs{\bLambda}{i,i} > 0$, and $\subs{\bLambda_{+}}{i,i}=1$, $\subs{\bLambda_b}{i,i}=0$ when $\subs{\bLambda}{i,i} = 0$. Two applications of \cref{lem:tbJ_prop}\ref{it:unitary} gives us $\tbJ_{\bx}(\bC, \bTheta) = \tbJ_\bx(\bC', \bLambda_b)$, where $\bC' = \bLambda_{+}^{-1/2}\bQ\T\bC$ is a block diagonal matrix since $\bQ$, $\bLambda_{+}$ and $\bC$ are all block diagonal matrices. Thus $(\bC', \bLambda_b) \in \calC$ and the claim follows.
\end{enumerate}
\end{IEEEproof}

\Cref{prop:normalized}\ref{it:normalized_noise} shows that the power of the additive noise can be normalized by the compression matrix when sanitizing the data. On the other hand, \cref{prop:normalized}\ref{it:only_noise} shows that adding only noise can approximate linear compression arbitrarily well in terms of the estimation error covariance, but this requires arbitrarily large noise power. Therefore, perturbation noise by itself cannot replace linear compression in a practical system. Without loss of generality, we set the compression matrix $\bC = \bI_N$ in the sequel as we can always normalize the noise by a compression matrix to obtain the same utility and privacy tradeoff.
We now rewrite the optimization problem \cref{opt:P0} as
\begin{align}\tag{P1} \label{opt:P1}
\begin{aligned}
	\max_{(\bI_N, \bTheta)\in \calC} \ & {u(\bI_N, \bTheta)}, \\
	{\rm s.t.}\ & p_i(\bI_N, \bTheta) \geq \epsilon_i, i = 1, \ldots, S.
\end{aligned}
\end{align}

Note that for \cref{opt:P1} to be feasible when $\bJ_0 \neq \bzero_L$, the privacy threshold $\epsilon_i$ should satisfy $\epsilon_i \leq {\epsilon_i}^{\max}$, where
\begin{align*}
{\epsilon_i}^{\max} = \trace{\bG_i \bJ_0^{-1} \bG_i\T} / \trace{\bG_i \bP_\bx \bG_i\T} - 1,
\end{align*}
for $\bG_i\ne\bzero$, and ${\epsilon_i}^{\max}=0$ if $\bG_i=\bzero$. This is because the prior information (as quantified by its \gls{FIM} $\bJ_0$) already leaks privacy and any sanitization cannot achieve a higher level of privacy than this. When $\bJ_0 = \bzero_L$, we have ${\epsilon_i}^{\max} = \infty$ if $\bG_i\ne\bzero$ and $0$ otherwise.

The following useful expressions can be derived by using the binomial inverse theorem and Woodbury matrix identity (see \cref{appendix:D} for details).
\begin{enumerate}
\item No prior information, i.e., $\bJ_0=\bzero_L$, we have
\begin{align}\label{eq:decomposed_eq_no_prior} 
\tbP_{\bx}(\bI_N, \bTheta) 
&= \bP_\bx + \bPsi \inv{\bI_N + \bTheta \bPhi} \bTheta \bPsi\T, 
\end{align}
where 
\begin{align}
\bPhi &= \bR^{-1} - \bR^{-1} \bH \bP_\bx \bH\T \bR^{-1}\label{eq:decomposed_eq_no_prior_Phi} \\
\intertext{is a degenerate matrix and} 
\bPsi &= \bP_\bx \bH\T \bR^{-1}. \label{eq:decomposed_eq_no_prior_Psi} 
\end{align}

\item With prior information, i.e., $\bJ_0 \succ \bzero_L$, we have
\begin{align} 
 \tbP_{\bx}\left( \bI_N, \bTheta\right) 
&= \bP_0 - \bP_0 \bH\T \inv{\bPhi^{-1} + \bTheta} \bH \bP_0 \label{eq:perturbed_CRLB_with_prior}\\  
&= \bP_\bx + \bPsi \inv{\bI_N + \bTheta \bPhi}\bTheta \bPsi\T, \label{eq:decomposed_eq_with_prior} 
\end{align}
where $\bP_0 = \bJ_0^{-1}$,
\begin{align}
\bPhi &= \inv{\bH\bP_0\bH\T + \bR}, \label{eq:decomposed_eq_with_prior_Phi}
\intertext{and} 
\bPsi &= \bP_0 \bH\T \bPhi. \label{eq:decomposed_eq_with_prior_Psi}
\end{align}
\end{enumerate}

By considering the observation model $\by=\bH\bx+\bn$, $\bPsi$ in \cref{eq:decomposed_eq_no_prior} is the same as the minimum mean square error estimation matrix. By regarding $\bP_0$ as the predicted estimate covariance, $\bPsi$ and $\bPhi^{-1}$ in \cref{eq:decomposed_eq_with_prior} are equivalent to the Kalman gain and innovation covariance, respectively. The equations \cref{eq:decomposed_eq_no_prior,eq:decomposed_eq_with_prior} under both the cases where prior information is available and unavailable have meaningful interpretations: the first term of both equations is the \gls{CRLB} without perturbation while the second term is the increased \gls{CRLB} caused by noise perturbation.

%---------------------------------------------------------------------------%
%                                    ASUP                                   %
%---------------------------------------------------------------------------%
\section{Arbitrarily Strong Privacy with Perfect Utility} \label{sect:asup}
In this section, we derive necessary and sufficient conditions for achieving perfect utility and arbitrarily strong privacy when prior information is available and unavailable, respectively.

\begin{Definition}\label{def:ASUP}
We say that arbitrarily strong utility-privacy tradeoff (ASUP) is achievable if for any non-negative $\epsilon_i < \epsilon_i^{\max}$, $i=1,\ldots,S$, there exists $(\bC,\bTheta)\in\calC$ such that $u(\bC,\bTheta) = 0$ and $p_i(\bC,\bTheta) \geq \epsilon_i$ for all $i=1,\ldots,S$.
\end{Definition}
While achieving perfect utility (by taking $\calT$ to be $(\bI_N,\bzero_N)$ and arbitrarily strong privacy (by taking $\calT$ to be $(\bzero_N,\bzero_N)$) are easy, achieving both at the same time is difficult or infeasible. Therefore, it is of interest to know when ASUP is achievable. In the following, we derive necessary and sufficient conditions for ASUP under both the cases where prior information is available and unavailable. Moreover, we provide a method to construct the noise covariance to achieve ASUP inside the proofs (if ASUP is feasible). We start with the following two preliminary lemmas.

% preliminary lemma
\begin{Lemma} \label{lemma:trace_matrix_inverse_equivalence}
For matrices $\bA \in \bbR^{M \times N}$, $\bB \in \bbS^N$ and a sequence of pairwise commuting matrices $(\bS_n)_{n\geq1}$ where each $\bS_n \subset \bbS_+^N$, we have
\begin{align*}
\lim_{n \to \infty}\trace{\bA \inv{\bS_n + \bB} \bA\T} = 0,
\end{align*}
if and only if
\begin{align*}
\lim_{n \to \infty} \trace{\bA \bS_n^{-1} \bA\T} = 0.
\end{align*}
\end{Lemma}
\begin{IEEEproof}
See \cref{sec:lemma_eqiv}.
\end{IEEEproof}

% preliminary lemma
\begin{Lemma} \label{lemma:vector_orthogonality}
For matrices $\bA \in \bbR^{M_1 \times N}$ and $\bB \in \bbR^{M_2 \times N}$ such that $M_1,M_2\leq N$, there exists $\bS \in \bbS^N$ such that $\trace{\bA\bS\bA\T} = 0$ and
\begin{align*}
\lim_{n \to \infty}&\trace{\bB \inv{n\bS+\frac{1}{n}\bI_N}\bB\T} = 0, 
\end{align*}
if and only if $\bA\bB\T = \bzero$.
\end{Lemma}
\begin{IEEEproof}
See \cref{sec:vector_orthogonality}.
\end{IEEEproof}
%
% Theorem ASUP (No Piror)
\subsection{ASUP without prior information}

In this subsection, we consider the case $\bJ_0=\bzero_L$.
\begin{Theorem} \label{thm:asup_no_prior}
Suppose $\bJ_0=\bzero_L$. Recall $\bPhi$ and $\bPsi$ as defined in \cref{eq:decomposed_eq_no_prior_Phi,eq:decomposed_eq_no_prior_Psi}, respectively. 
For $i\in\{1,\ldots,S\}$, let $\bXi_i=\begin{bsmallmatrix} \bU \subs{\bPsi}{:,{\calS}_i} \\ \subs{\bPhi}{\calS_i,\calS_i} \end{bsmallmatrix}$.
There exists $\bTheta\in \bbS^N$ such that ASUP is achievable if and only if for each $\bG_j\neq\bzero$ where $j\in\{1,\ldots,S\}$, there exists at least an agent $i\in\{1,\ldots,S\}$ such that
\begin{enumerate}[label={(\roman*)},ref={\roman*}]
\item \label[condition]{it:J0_cond1} $N_i > \rank(\bXi_i)$, and 
\item \label[condition]{it:J0_cond2} there does not exist any matrix $\bP$ such that $\bP\bXi_i=\bG_j\subs{\bPsi}{:,{\calS}_i}$.
\end{enumerate}
\end{Theorem}
\begin{IEEEproof} Note that \cref{it:J0_cond1} is equivalent to 
\begin{align} \label{null_space:J0_cond1}
\Null{\subs{\bPhi}{\calS_i,\calS_i}} 
\cap \Null{\bU \subs{\bPsi}{:,{\calS}_i}} 
\neq \{\bzero\},
\end{align}
and \cref{it:J0_cond2} is equivalent to 
\begin{align} \label{null_space:J0_cond2}
\begin{split}
&\left(\Null{\subs{\bPhi}{\calS_i,\calS_i}} 
\cap \Null{\bU\subs{\bPsi}{:,{\calS}_i}}\right) \\
&\nsubseteq\Null{\bG_j\subs{\bPsi}{:,{\calS}_i}}.
\end{split}
\end{align}

We first show the necessity of \cref{it:J0_cond1,it:J0_cond2}. From \cref{eq:decomposed_eq_no_prior,eq:Utility_Loss}, perfect utility $u(\bI_N, \bTheta) = 0$ is obtained only if
\begin{align*}
\trace{\bU\bPsi \inv{\bI_N + \bTheta \bPhi} \bTheta \bPsi\T\bU\T} = 0.
\end{align*}
Since the \gls{LHS} of the above equation is a continuous function \gls{wrt} $\bTheta$, the above equation can be expressed as
\begin{align*}
\lim_{n \to \infty} \trace{\bU\bPsi \inv{\inv{\bTheta + \frac{1}{n}\bI_N} + \bPhi} \bPsi\T\bU\T} = 0.
\end{align*}
From \cref{lemma:trace_matrix_inverse_equivalence}, this is equivalent to
\begin{align*}
\lim_{n \to \infty} \trace{\bU\bPsi \left(\bTheta + \frac{1}{n}\bI_N\right) \bPsi\T\bU\T} = 0,
\end{align*}
or 
\begin{align*}
\trace{\bU \bPsi \bTheta \bPsi\T\bU\T} = 0.
\end{align*}
Therefore, $u(\bI_N, \bTheta) = 0$ only if $\bU\bPsi\bTheta = \bzero$. Following that, $\bU\subs{\bPsi}{:,{\calS}_k}\bTheta_k=\bzero$, for all $k=1,\ldots,S$ since $\bTheta=\diag\left(\bTheta_1,\ldots,\bTheta_S\right)$. From the rank-nullity theorem, this implies that $\bTheta_k=\bzero_{N_k}$ for an agent $k$ with $\rank\left(\bU\subs{\bPsi}{:,{\calS}_k}\right)=N_k$. Since $\bG_j\ne\bzero$, to achieve any level of privacy, there must exist at least an agent $i$ with $\rank\left(\bU\subs{\bPsi}{:,{\calS}_i}\right)<N_i$ with $\bU \subs{\bPsi}{:,{\calS}_i}$ having a non-trivial null space. Let $\bTheta$ be such that $\bTheta_i\ne\bzero_{N_i}$ and
\begin{align} \label{eq:agent_k_for_u=0}
\bU\subs{\bPsi}{:,{\calS}_i} \bTheta_i =\bzero.
\end{align}

Since $p_j(\bI_N, \bTheta)$ is a non-decreasing function \gls{wrt} $\bTheta$, arbitrarily strong privacy for the private parameter $\bg_j$ is achieved only if
\begin{align}
\lim_{n \to \infty} p_j(\bI_N, n\bTheta) =\infty. \label{p_infty}
\end{align}
From \cref{eq:decomposed_eq_no_prior,eq:Privacy_Gain}, the \gls{LHS} of the above statement \cref{p_infty} is equivalent to
\begin{align*} 
&\lim_{n\to\infty} \trace{\bG_j\bPsi\inv{\bI_N+n\bTheta\bPhi} n\bTheta\bPsi\T\bG_j\T} \nn 
&=\lim_{n\to\infty}\trace{\bG_j\bPsi\inv{\inv{n\bTheta+\frac{1}{n}\bI_N}+\bPhi}\bPsi\T\bG_j\T} \nn
&= \lim_{n\to\infty}\sum_{k=1}^N \alpha_k(n,j)\lambda_k(n),
\end{align*}
where $\bT(n)=\inv{\inv{n\bTheta+\frac{1}{n}\bI_N}+\bPhi}$, $\lambda_k(n)$ is the $k$-th eigenvalue of $\bT(n)$ with $\bv_k(n)$ being the corresponding unit eigenvector, and $\alpha_k(n,j) = \norm{\bG_j\bPsi\bv_k(n)}_2^2$. Therefore, \cref{p_infty} holds only if there exists an index $k_j$ such that $\lim_{n\to\infty}\lambda_{k_j}(n)=\infty$ with corresponding eigenvectors $\bv_{k_j}(n) \notin \Null{\bG_j\bPsi}$. Since $\bG_j\bPsi$ has finite rank, by passing to a subsequence if necessary, there exists a unit vector $\bw_j \notin \Null{\bG_j\bPsi}$ such that
\begin{align*} 
\lim_{n \to \infty}\bw_j\T\inv{\bT(n)}\bw_j = 0.
\end{align*}
We then have
\begin{align}\label{eq:p>0_cond1}
\begin{aligned} 
&\lim_{n \to \infty} \bw_j\T \inv{n\bTheta + \frac{1}{n} \bI_N} \bw_j = 0, \text{ and}\\ 
&\bw_j\T\bPhi\bw_j = 0.
\end{aligned}
\end{align}
Since $\bTheta_i\ne \bzero_{N_i}$, \cref{eq:p>0_cond1} holds only if $\bw'_j=[\bw_j]_{\calS_i}\notin\Null{\bG_j\subs{\bPsi}{:,{\calS}_i}}$ and $\bw'_j\ne\bzero$, such that
\begin{align} 
&\lim_{n \to \infty} {\bw'_j}\T\inv{n\bTheta_i + \frac{1}{n}\bI_{N_i}} \bw'_j = 0, \text{ and}\label{eq:p>0_cond2}\\ 
&\subs{\bPhi}{\calS_i,\calS_i} \bw'_j = 0.\label{eq:p>0_cond3}
\end{align}
From \cref{lemma:vector_orthogonality}, both \cref{eq:agent_k_for_u=0} and \cref{eq:p>0_cond2} hold only if $\bw'_j \in \Null{\bU \subs{\bPsi}{:,{\calS}_i}}$. Therefore, \cref{it:J0_cond1,it:J0_cond2} are necessary for ASUP.

We next prove sufficiency by constructing a $\bTheta$ that yields ASUP. Let a unit vector $\bv'$ satisfy
\begin{enumerate}[(a)]
\item $\bv' \in \Null{\subs{\bPhi}{\calS_i,\calS_i}}$, and
\item $\bv' \in \Null{\bU \subs{\bPsi}{:,{\calS}_i}}$, and
\item $\bv' \notin\Null{\bG_j\subs{\bPsi}{:,{\calS}_i}}$.
\end{enumerate}
The existence of $\bv'$ is guaranteed by condition \ref{it:J0_cond1} and \ref{it:J0_cond2}. Set all the eigenvalues of $\bTheta_i$ to $0$ except for one eigenvalue $\lambda'>0$ associated with a unit eigenvector $\bv'$. Let $\widetilde{\bTheta}_j=\diag(\ldots,\bzero_{N_{i-1}},\bTheta_i,,\bzero_{N_{i+1}}\ldots)$. By reversing the arguments used in the necessity proof, it can be verified that for any $\epsilon_j>0$, there exists a large enough $\lambda'$ such that $u(\bI_N,\widetilde{\bTheta}_j)=0$ and $p_j(\bI_N,\widetilde{\bTheta}_j)\geq\epsilon_j$. Let $\bTheta$ be the sum of $\widetilde{\bTheta}_j$ over $j$ and such $\bTheta$ leads to ASUP. The proof is now complete.
\end{IEEEproof}

\begin{algorithm}[!tb]
\caption{ASUP sanitization algorithm ($\bJ_0=\bzero_L$).}\label{algo:asup_no_prior}
\begin{algorithmic}[1]
\REQUIRE{For each agent $i$, $\bPsi$ in \cref{eq:decomposed_eq_no_prior_Psi}, $p_i(\bI_N, \bTheta)$ in \cref{eq:privacy_function}, and $\bXi_i$ as defined in \cref{thm:asup_no_prior}. Conditions of \cref{thm:asup_no_prior} are satisfied.}
\FOR{each agent $i=1,\ldots,S$}
\STATE{Choose an orthonormal basis $\bW_i$ of $\Null{\bXi_i}$.}
\STATE{Choose a row vector $\bv_i$ from $\bW_i$ such that $\bv_i \notin \Null{\bG_i\subs{\bPsi}{:,{\calS}_i}}$.}
\STATE{Choose a matrix $\bX_i$ such that $\bV_i=\left[\bv_i\T,\bX_i\right]$ is unitary. Set $\bTheta_i=\bV_i\diag(\lambda_i,\bzero_{N_i-1})\bV_i\T$ with $\lambda_i>0$ and $\widetilde{\bTheta}_i=\diag(\ldots,\bzero_{N_{i-1}},\bTheta_i,\bzero_{N_{i+1}},\ldots)$. Choose $\lambda_i$ to be large enough such that $p_i(\bI_N,\widetilde{\bTheta}_i)\geq\epsilon_i$.}
\STATE{Agent $i$ generates a noise $\bxi_i \sim \N{\bzero}{\bTheta_i}$ and adds this to its measurement $\by_i$.}
\ENDFOR
\end{algorithmic}
\end{algorithm}

%\cref{thm:asup_no_prior} implies that ASUP is achievable when $\bJ_0=\bzero_L$ as long as one agent in the network satisfies certain conditions. 
As an illustration, consider the special case $\bH \in \bbR^{L \times L}$, where $\bPsi$ turns out to be $\bH^{-1}$ and $\bPhi$ becomes $\bzero_N$. Then from \cref{eq:decomposed_eq_no_prior}, the perturbed \gls{CRLB} for estimating $\bx$ can be written as
\begin{align*}
\tbP_{\bx}(\bI_N, \bTheta)
&= \bP_\bx + \bH^{-1} \bTheta \inv{\bH\T} \\
&= \bP_\bx + \sum_{i=1}^S \subs{\bH^{-1}}{:,\calS_i} \bTheta_i \subs{\inv{\bH\T}}{:,\calS_i}\T,
\end{align*}
which is linear in the perturbation noise covariance $\bTheta_i$ of each agent $i$. Recall that $\bu=\bU\bx$ and $\bg_i=\bG_i\bx$. It can be seen that if there is one row in $\bU\subs{\bH^{-1}}{:,\calS_i}$ not in the row space of $\bG_i\subs{\bH^{-1}}{:,\calS_i}$, agent $i$ can use arbitrarily large noise power in the corresponding row of $\bTheta_i$ so that the estimation error of $\bg_i$ becomes arbitrarily large while the estimation error of $\bu$ remains unchanged. \cref{thm:asup_no_prior} generalizes this result to any $\bH$. Based on \cref{thm:asup_no_prior}, we can implement a decentralized sanitization scheme as illustrated in \cref{algo:asup_no_prior} if we have $N_i>\rank(\bXi_i)$ and $\Null{\bXi_i}\nsubseteq\Null{\bG_i\subs{\bPsi}{:,{\calS}_i}}$ for each agent $i=1,\ldots,S$.

% Theorem ASUP (With Piror)
\subsection{ASUP with prior information}
In this subsection, we consider the case $\bJ_0 \succ \bzero_L$.

\begin{Theorem} \label{thm:asup_with_prior}
Suppose $\bJ_0 \succ \bzero_L$. ASUP is achievable if and only if
\begin{align}
\bU \subs{\bPsi}{:,\calS_i} {\bH_i} \bP_0 \bG\T  =\bzero,\label[condition]{it:J0>0_cond}
\end{align}
for every agent $i=1,\ldots,S$, where $\bG=\left[\bG_1\T,\ldots,\bG_S\T\right]\T\in\Real^{G\times L}$.
\end{Theorem}
\begin{IEEEproof}
From \cref{eq:decomposed_eq_with_prior}, perfect utility $ u(\bI_N, \bTheta) = 0$ requires
\begin{align*}
\Tr\left(\bU\bPsi \inv{\bI_N + \bTheta \bPhi}\bTheta \bPsi\T \bU\T\right) = 0.
\end{align*}
Applying \cref{lemma:trace_matrix_inverse_equivalence} and the same argument in the proof of \cref{thm:asup_no_prior}, the above statement holds if and only if
\begin{align} \label{eqn:Jn0_cond1}
\trace{\bU \bPsi \bTheta \bPsi\T\bU\T} = 0.
\end{align}
Arbitrarily strong privacy requires the existence of $\bTheta$ such that 
\begin{align*}
\lim_{n\to\infty}p_i(\bI_N,n\bTheta)=\epsilon_i^{\max},\ \forall\ i=1,\ldots,S.
\end{align*} 
From \cref{eq:perturbed_CRLB_with_prior}, the above equation can be written as
\begin{align*} 
\lim_{n\to\infty} \Tr\left(\bG\bP_0\bH\T\inv{\bPhi^{-1}+n\bTheta} \bH\bP_0\bG\T\right) = 0.
\end{align*}
Since the \gls{LHS} of the above statement is a continuous function \gls{wrt} $\bTheta$, it is equivalent to
\begin{align*}
\lim_{n\to\infty} \Tr\left(\bG\bP_0\bH\T\inv{\bPhi^{-1} + n\bTheta + \frac{1}{n}\bI_N}\bH\bP_0\bG\T\right) = 0.
\end{align*}
From \cref{lemma:trace_matrix_inverse_equivalence}, we have equivalently,
\begin{align} \label{eqn:Jn0_cond2}
\lim_{n\to\infty}\Tr\left(\bG\bP_0\bH\T\inv{n\bTheta + \frac{1}{n}\bI_N} \bH\bP_0\bG\T\right) = 0.
\end{align}
Rewriting the perfect utility condition \cref{eqn:Jn0_cond1} and the arbitrarily strong privacy condition \cref{eqn:Jn0_cond2} in terms of $\bTheta_i$, $i=1,\ldots,S$, yields
\begin{align*} 
&\sum_{i=1}^S\Tr\left(\bU \subs{\bPsi}{:,\calS_i} \bTheta_i \subs{\bPsi}{:,\calS_i}\T \bU\T\right) = 0, \\
&\lim_{n \to \infty} \sum_{i=1}^S\Tr\left(\bG\bP_0 {\bH_i\T} \inv{n\bTheta_i + \frac{1}{n}\bI_{N_i}} {\bH_i} \bP_0 \bG\T\right) = 0.
\end{align*}
The proof of necessity now follows by applying \cref{lemma:vector_orthogonality} to the above statements. To show sufficiency, we can follow the method in Appendix~\ref{sec:vector_orthogonality} to construct $\bTheta$ as illustrated in \cref{algo:asup_with_prior}, which guarantees $p_i(\bI_N,\bTheta)\geq\epsilon_i$, for any given $\epsilon_i<\epsilon_i^{\max}$, $i=1,\ldots,S$. The theorem is now proved.
\end{IEEEproof}
\begin{algorithm}[!tb]
\caption{ASUP sanitization algorithm ($\bJ_0\succ\bzero_L$).}\label{algo:asup_with_prior}
\begin{algorithmic}[1]
\REQUIRE{For each agent $i$, $\bPsi$ in \cref{eq:decomposed_eq_with_prior_Psi} and $\bXi_i=\bG\bJ_0^{-1}\bH_i\T$ with $\bG$ defined in \cref{thm:asup_with_prior}. Condition in \cref{thm:asup_with_prior} is satisfied.}
\FOR{each agent $i=1,\ldots,S$}
\STATE{Choose an orthonormal basis $\bW_i\in\bbR^{W_i\times N_i}$ of the row space of $\bXi_i$.}
\STATE{Choose an orthonormal basis $\bY_i$ of the row space of $\bU \subs{\bPsi}{:,\calS_i}$.}
\STATE{Choose a matrix $\bX_i$ such that $\bV_i\T=\left[\bW_i\T,\bY_i\T,\bX_i\T\right]$ is unitary. Choose a diagonal matrix $\bLambda_i\succeq\bzero_{W_i}$ such that $\Tr(\bXi_i\bW_i\T\bLambda_i^{-1}\bW_i\bXi_i\T)\leq \epsilon_i^{\max}-\epsilon_i$. 
Let $\bTheta_i=\bV_i\T\diag(\bLambda_i,\bzero_{N_i-W_i})\bV_i$.}%
\STATE{Agent $i$ generates a noise $\bxi_i \sim \N{\bzero}{\bTheta_i}$ and adds this to its measurement $\by_i$.}
\ENDFOR
\end{algorithmic}
\end{algorithm}
Note that $\bPsi\bH\bP_0$ is the reduced estimate covariance for $\bx$ \gls{wrt} the prior estimate covariance $\bP_0$ after agents observe $\by$. We may consider $\subs{\bPsi}{:,\calS_i} \bH_i \bP_0$ in \cref{it:J0>0_cond} as the reduced estimate covariance for agent $i$ to observe $\by_i$. \cref{thm:asup_with_prior} gives the condition for each agent $i=1,\ldots,S$, to be able to decrease this reduced estimate covariance to its minimum for the private parameter $\bg_i=\bG_i\bx$ without affecting the reduced estimate covariance for the public parameter $\bu=\bU\bx$.

We note that the privacy function is bounded when $\bJ_0 \succ \bzero_L$ because $\bJ_0\succ \bzero_L$ provides prior information about every agent's private parameter. The achievable privacy is the cumulative result of the noise perturbation applied by each agent. Since the privacy contributed by each agent is bounded when $\bJ_0 \succ \bzero_L$, the maximum privacy is only obtained when all the agents offer their maximum privacy. This is different from the case $\bJ_0=\bzero_L$, where the privacy each agent can contribute is unbounded. Therefore, we can rely on one agent to provide arbitrarily strong privacy in \cref{thm:asup_no_prior}. This explains why ASUP for $\bJ_0=\bzero_L$ requires only one agent while ASUP for $\bJ_0 \succ \bzero_L$ requires every agent. In the case where only a set of agents satisfy the condition given in \cref{thm:asup_with_prior} when $\bJ_0 \succ \bzero_L$, we still can apply the sanitization method in \cref{algo:asup_with_prior} to these nodes. However, as a consequence, we will obtain perfect utility but not arbitrarily strong privacy. In what follows we give an example in which \cref{it:J0>0_cond} in \cref{thm:asup_with_prior} holds.

\begin{Example}
Suppose $\bH_i= \bH_0 \in \Real^{(N/S)\times L}$, for all $i=1,\ldots,S$, and the measurement noise is white, i.e., $\bR = \sigma^2\bI_N$. In this case, we can verify that for some $\bPsi_0\in\Real^{N\times (N/S)}$, $\subs{\bPsi}{:,\calS_i}=\bPsi_0$, for all $i=1,\ldots,S$. When $\bU\bPsi_0\bH_0\bP_0\bG\T=\bzero$, \cref{it:J0>0_cond} is satisfied. Note that $\bPsi_0\bH_0\bP_0$ is the reduced estimate covariance of $\bx$ after agent $i$ observes $\by_i$. Following that, $\bU\bPsi_0\bH_0\bP_0\bG\T$ can be interpreted as the reduced estimate cross-covariance between the public parameter $\bu$ and the private parameters $\bg_1,\ldots,\bg_S$, which need to be uncorrelated to make ASUP feasible.
\end{Example}

\section{Maximum privacy under perfect utility} \label{sect:max_privacy}
In this section, we formulate the problem of achieving maximal privacy under perfect utility (i.e., $u(\bC,\bTheta)=0$) as a linear matrix inequality optimization. When discussing ASUP, we allow the power of the perturbation noise to be arbitrarily large so that the privacy function can exceed arbitrarily large privacy thresholds. In most practical applications, there is a power constraint on the perturbation noise that can be added (if we normalize the noise as discussed immediately before \cref{opt:P1}, this translates into a constraint on the dynamic range of the compression mechanism.). 

We consider the following optimization:
\begin{align}\label{opt:P2}
\begin{aligned}
\max_{\left(\bI_N, \bTheta\right) \in \calC} \ & \sum_{i=1}^S {p_i(\bI_N, \bTheta)}, \\
\st\ &u(\bI_N, \bTheta) = 0, \\
&\trace{\bTheta_i} \leq \delta_i,\ \text{for } i=1,\ldots,S,
\end{aligned}
\end{align}
where $\delta_i>0$ is the maximum total power available to agent $i$. In the following, we show how to reformulate \cref{opt:P2} as a convex optimization problem with linear matrix inequality (LMI) constraints.

Recall from \cref{thm:asup_no_prior,thm:asup_with_prior} that $u(\bI_N,\bTheta)=0$ under both the cases $\bJ_0=\bzero_L$ and $\bJ_0 \succ \bzero_L$ is achievable if and only if
\begin{align*}
\trace{\bU \bPsi \bTheta \bPsi\T \bU\T} = 0.
\end{align*}
By introducing an auxiliary variable $\bZ \in \bbS^{L}$, \cref{opt:P2} can be cast as 
\begin{align} \label{eq:constraint_privacy}
\begin{aligned}
	\max_{\left(\bI_N, \bTheta\right) \in \calC, \bZ \in \bbS^{L}} \ & \sum_{i=1}^S \bG_i \bZ \bG_i\T, \\ 
	{\rm s.t.}\ &\bZ \preceq \tbP_{\bx}(\bI_N, \bTheta), \\ 
	& \trace{\bU \bPsi \bTheta \bPsi\T \bU\T} = 0, \\
	&\diag\parens*{(\trace{\bTheta_i})_{i=1}^S} \leq \diag\parens*{(\delta_i)_{i=1}^S}.
\end{aligned}
\end{align}
Note that the maximization will force $\bZ$ to achieve its upper bound. 

In the case $\bJ_0=\bzero_L$, we need the following form of the perturbed \gls{CRLB} before transforming the first constraint of \cref{eq:constraint_privacy} into a LMI. Let the column vectors of $\overline{\bH}$ be a basis of $\Null{\bH\T}$ and let $\bK=\left[\bK_1\T,\bK_2\T\right]\T=\left[\bH,\overline{\bH}\right]^{-1}$ with $\bK_1 \in \bbR^{L \times N}$. We obtain from \cref{eq:perturbed_CRLB}
\begin{align*}
\tbP_\bx(\bI_N,\bTheta)
&= \inv{\bH\T\inv{\bR+\bTheta}\bH} \\
&= \inv{\subs{\inv{\bK\T}\inv{\bR+\bTheta}\bK^{-1}}{\calL,\calL}} \\
&= \inv{\subs{\inv{\bK(\bR+\bTheta)\bK\T}}{\calL,\calL}} \\
&= \bK_1\tbTheta\bK_1\T - \bK_1\tbTheta\bK_2\T\inv{\bK_2\tbTheta\bK_2\T}\bK_2\tbTheta\bK_1\T,
\end{align*}
where $\tbTheta = \bR + \bTheta$, $\calL = \{1, \ldots, L\}$ and the last equation is a consequence of the block matrix inversion.
Substituting the above expression into the first constraint of \cref{eq:constraint_privacy} and using the Schur complement, the constraint can be cast as the following LMI:
\begin{align*}
\begin{bmatrix}
\bK_1(\bTheta+\bR)\bK_1\T & \bK_1(\bTheta+\bR)\bK_2\T \\ 
\bK_2(\bTheta+\bR)\bK_1\T & \bK_2(\bTheta+\bR)\bK_2\T
\end{bmatrix}
-
\begin{bmatrix}
\bZ & \bzero \\ 
\bzero & \bzero_{N-L}
\end{bmatrix}
\succeq \bzero_N.
\end{align*}
Similarly, in the case $\bJ_0 \succ \bzero_L$, by substituting \cref{eq:perturbed_CRLB_with_prior} into the first constraint of \cref{eq:constraint_privacy}, it can be linearized as
\begin{align*}
\begin{bmatrix}
\bP_0 - \bZ & \bP_0 \bH\T \\ \bH \bP_0 & \bPhi^{-1} + \bTheta
\end{bmatrix}
\succeq \bzero_{N+L}.
\end{align*}
The above formulations can then be solved by using standard semi-definite programming techniques \cite{BoyVan:B04}.

Together with the results in \cref{thm:asup_no_prior,thm:asup_with_prior}, one may proceed to design the sanitization mechanism at each agent as follows. One first checks if the necessary and sufficient conditions for achieving ASUP are satisfied. If so and if there is no power constraint, we can use the steps outlined in \cref{algo:asup_no_prior,algo:asup_with_prior} to choose the sanitization mappings. If there is a power constraint, we solve \cref{eq:constraint_privacy}. On the other hand, if ASUP is not achievable, we propose an alternating optimization procedure to solve \cref{opt:P1} in the following section. 

\section{Alternating Optimization} \label{sect:sequential_optimization}
In this section, we propose an algorithm to solve \cref{opt:P1} for both the cases where information is available and unavailable. Because $\bTheta$ is restricted to be a block diagonal matrix and the utility and privacy functions are non-linear \gls{wrt} $\bTheta$, it is difficult to directly solve the optimization problem \cref{opt:P1} using standard optimization tools. We propose to optimize \cref{opt:P1} \gls{wrt} $\bTheta_1, \ldots, \bTheta_S$, sequentially and in an alternating fashion, instead of $\bTheta$. We show that optimizing \cref{opt:P1} over $\bTheta_i$ is a convex problem, which can be solved by semi-definite programming \cite{BoyVan:B04}.

To proceed, we need to do a trivial relaxation by assuming $\bTheta$ is invertible. Then both the perturbed \gls{CRLB}s when $\bJ_0=\bzero_L$ and $\bJ_0 \succ \bzero_L$ in \cref{eq:decomposed_eq_no_prior,eq:decomposed_eq_with_prior} can be written as
\begin{align} \label{eq:seq_opt_crlb}
\tbP_{\bx}\left( \bI_N, \bTheta\right) = \bP_\bx + \bPsi \inv{\bTheta^{-1} + \bPhi} \bPsi\T,
\end{align}
where $\bPsi$ and $\bPhi$ refer to their respective definitions when prior information is available or unavailable. Therefore, the algorithm to be proposed when $\bJ_0=\bzero_L$ and $\bJ_0 \succ \bzero_L$ can be described as one. For sequential optimization, only $\bTheta_i$ is updated at the $i$-th sequence while the rest of the parameters are fixed. By doing this, the utility and privacy tradeoff can be cast as a standard semi-definite programming problem \gls{wrt} $\bTheta_i$. Denote $\overline{\bTheta} = \bTheta^{-1}$ and for $i =1,\ldots,S$, let $\overline{\bTheta}_i = \bTheta_i^{-1}$ and
\begin{align*}
\overline{\bTheta}_{\bslash{i}} = \diag(\overline{\bTheta}_{1}, \ldots, \overline{\bTheta}_{i-1}, \overline{\bTheta}_{i+1}, \ldots, \bTheta_{S}).
\end{align*}
We can permute the rows of $\overline{\bTheta}$ to obtain $\diag(\overline{\bTheta}_{i}, \overline{\bTheta}_{\bslash{i}})$. In the same way, we permute the rows of $\bH$ and $\bR$ to obtain 
\begin{align*}
&\left[\bH_i\T, \bH_1\T, \ldots, \bH_{i-1}\T, \bH_{i+1}\T, \ldots, \bH_{S}\T \right]\T, \\
&\diag(\bR_i, \bR_1, \ldots, \bR_{i-1}, \bR_{i+1}, \ldots, \bR_S),
\end{align*}
respectively. Substitute the above expressions for $\bH$ and $\bR$, respectively, in $\bPhi$ and $\bPsi$ to obtain $\bPhi_i$ and $\bPsi_i$. Treating $\overline{\bTheta}_{\bslash{i}}$ as a constant, the perturbed \gls{CRLB} \cref{eq:seq_opt_crlb} can be written as a function of $\overline{\bTheta}_i$:
\begin{align*}
\overline{\bP}_{\bx} (\overline{\bTheta}_{i}) = \bP_\bx + \bPsi_i \inv{\diag(\overline{\bTheta}_{i}, \overline{\bTheta}_{\bslash{i}}) + \bPhi_i} {\bPsi_i}\T.
\end{align*}
Note the above expression of the \gls{CRLB} is equal to the \gls{CRLB} in \cref{eq:seq_opt_crlb} since the permutation process keeps the \gls{CRLB} unchanged.

Consequently, the utility and $j$-th privacy function for $j=1,\ldots,S$, \gls{wrt} $\overline{\bTheta}_i$ can be expressed as (see {Appendix~\ref{appendix:E}} for details)
\begin{align}
&u(\overline{\bTheta}_{i}) 
= - \trace{\bGamma_u \inv{\overline{\bTheta}_{i} + \bOmega}} / \trace{\bU\bP_\bx\bU\T} - \bDelta_u,\label{alt_u}\\
&{p_j}(\overline{\bTheta}_{i}) 
=  \trace{\bGamma_{p_j} \inv{\overline{\bTheta}_{i} + \bOmega}} / \trace{\bG_j\bP_\bx\bG_j\T} + \bDelta_{p_j}, \label{alt_p}
\end{align}
where $\bOmega, \bDelta_{p_j}, \bDelta_u, \bGamma_u, \bGamma_p$ are constants independent of $\overline{\bTheta}_i$ given by
\begin{align*}
&\calN_i 
= \{1, \ldots, N_i\},\ \calN_{\bslash i} = \{N_i+1, \ldots, N\},\\
&\bDelta_{p_j}
= \trace{\bG_j \subs{\bPsi_i}{:,\calN_{\bslash{i}}} \bT \subs{\bPsi_i}{:,\calN_{\bslash{i}}}\T \bG_j\T} / \trace{\bG_j\bP_\bx\bG_j\T},\\
&\bDelta_u
= \trace{\bU \subs{\bPsi_i}{:,\calN_{\bslash{i}}} \bT \subs{\bPsi_i}{:,\calN_{\bslash{i}}}\T \bU\T} / \trace{\bU\bP_\bx\bU\T}, \\
&\bGamma_u 
= \subs{\bPsi_i}{:,\calN_i}\T\bU\T\bU\subs{\bPsi_i}{:,\calN_i} \\
&\quad\quad -\subs{\bPsi_i}{:,\calN_i}\T\bU\T\bU\subs{\bPsi_i}{:,\calN_{\bslash{i}}}\bT \subs{\bPhi_i}{\calN_{\bslash i},\calN_i} \\
&\quad\quad -\subs{\bPhi_i}{\calN_i,\calN_{\bslash i}} \bT\subs{\bPsi_i}{:,\calN_{\bslash{i}}}\T\bU\T\bU\subs{\bPsi_i}{:,\calN_i} \\
&\quad\quad +\subs{\bPhi_i}{\calN_i,\calN_{\bslash i}} \bT \subs{\bPsi_i}{:,\calN_{\bslash{i}}}\T \bU\T\bU\subs{\bPsi_i}{:,\calN_{\bslash{i}}} \bT \subs{\bPhi_i}{\calN_{\bslash i},\calN_i}, \\
&\bGamma_{p_j} 
= \subs{\bPsi_i}{:,\calN_i}\T\bG_j\T\bG_j\subs{\bPsi_i}{:,\calN_i} \\
&\quad\quad -\subs{\bPsi_i}{:,\calN_i}\T\bG_j\T\bG_j\subs{\bPsi_i}{:,\calN_{\bslash{i}}}\bT \subs{\bPhi_i}{\calN_{\bslash i},\calN_i} \\
&\quad\quad -\subs{\bPhi_i}{\calN_i,\calN_{\bslash i}} \bT\subs{\bPsi_i}{:,\calN_{\bslash{i}}}\T\bG_j\T\bG_j\subs{\bPsi_i}{:,\calN_i} \\ 
&\quad\quad +\subs{\bPhi_i}{\calN_i,\calN_{\bslash i}} \bT \subs{\bPsi_i}{:,\calN_{\bslash{i}}}\T \bG_j\T\bG_j\subs{\bPsi_i}{:,\calN_{\bslash{i}}} \bT \subs{\bPhi_i}{\calN_{\bslash i},\calN_i}, \\
&\bOmega 
= \subs{\bPhi_i}{\calN_i,\calN_i} - \subs{\bPhi_i}{\calN_i,\calN_{\bslash i}} \bT \subs{\bPhi_i}{\calN_{\bslash i},\calN_i}, \\
&\bT 
= \inv{ \overline{\bTheta}_{\bslash{i}} + \subs{\bPhi_i}{\calN_{\bslash i},\calN_{\bslash i}} }.
\end{align*}
At the $i$-th sequence, we treat $\overline{\bTheta}_{\bslash{i}}$ as a constant and optimize \cref{opt:P1} over $\overline{\bTheta}_{i}$. This optimization problem can be formulated as
\begin{align*}%\tag{P2} \label{opt:P2}
\begin{split}
	\max_{\overline{\bTheta}_{i} \succeq \bzero_{N_i}} \ & u(\overline{\bTheta}_{i}), \\
	{\rm s.t.}\ & {p_j}(\overline{\bTheta}_{i}) \geq \epsilon_j, \forall\ j = 1,\ldots,S,
\end{split}
\end{align*}
which can be re-cast as a semidefinite programming problem by introducing $\bZ = \inv{\overline{\bTheta}_{i} + \bOmega}$ to obtain:
\begin{align*}\tag{P2} \label{opt:P3}
\begin{split}
	\min_{\bZ} \ & \trace{\bGamma_u \bZ}, \\
	{\rm s.t.}\ & \trace{\bGamma_{p_j} \bZ} \geq \epsilon_j', \forall\ j = 1, \ldots, S, \\
				& \bzero_{N_i} \preceq \bZ \preceq \bOmega^{-1},
\end{split}
\end{align*}
where $\epsilon_j' = \left(\epsilon_j - \bDelta_{p_j}\right) \times \trace{\bG_j\bP_\bx\bG_j\T} $. 

We propose to optimize \cref{opt:P1} over $\overline{\bTheta}_{1}, \ldots, \overline{\bTheta}_{S}$ sequentially over multiple iterations. The algorithm is summarized in \cref{algo:sequential_opt}.
\begin{algorithm}[!htb] 
\caption{Alternating optimization}\label{algo:sequential_opt}
\textbf{Initialize:} $\overline{\bTheta}_{i}^0 = \bzero_{N_i}, i = 1, \ldots, S$ at iteration $0$. \\
\textbf{Output:} $\overline{\bTheta}_{i}^k, i=1,\ldots,S$ at iteration $k$.

\begin{algorithmic}[1]
\WHILE {$k\leq$ \textbf{Max. number of iterations}}
\FOR {$i = 1$ \TO $S$}
\STATE Compute $\epsilon_j', \bGamma_u$ and $\bGamma_{p_j}$ for all $j=1, \ldots, S$.
\STATE {Solve \cref{opt:P3} to obtain $\overline{\bTheta}_{i}^{k}$}.
\STATE Use $\overline{\bTheta}_{i}^{k}$ to update $\overline{\bTheta}_{\bslash{i+1}}^k, \ldots, \overline{\bTheta}_{\bslash{S}}^k$.
\ENDFOR
\ENDWHILE
\end{algorithmic}
\end{algorithm}
%

%---------------------------------------------------------------------------%
%               Sect:  Numerical result and discussion                      %
%---------------------------------------------------------------------------%

\section{Numerical results} \label{sect:simu_discussion}
In the section, we investigate the impact of different parameters on the maximum privacy attainable under perfect utility discussed in \cref{sect:asup} and verify the performance of \cref{algo:sequential_opt} proposed in \cref{sect:sequential_optimization}. The settings used for the simulations in this section are given as follows unless otherwise stated:
\begin{itemize}
\item The total number of agents' observations, $N=72$.
\item The number of agent $i$'s observations, $N_i=N/S$ for $i=1,\ldots,S$.
\item The dimension of the hidden parameter $\bx$, $L=12$.
\item The measurement noise covariance $\bR=\bA\bA\T$, where the entries of $\bA\in\bbR^{72\times 72}$ are independent samples drawn from $\Unif{-0.5,0.5}$.
\item The entries of the observation model matrix $\bH\in\bbR^{72\times 12}$ are independent samples drawn from $\Unif{-0.5,0.5}$.
\item The prior information $\bP_0=\bA\bA\T$, where the entries of $\bA\in\bbR^{12\times 12}$ are independent samples drawn from $\Unif{-10,10}$.
\item The entries of the projection matrix of the public variable $\bU\in\bbR^{U\times 12}$ are independent samples drawn from $\Unif{-0.5,0.5}$.
\item The projection matrix of the $i$-th private variable, $\bG_i=\bG\in\bbR^{3\times 12}$, where the entries of $\bG$ are independent samples drawn from $\Unif{-0.5,0.5}$.
\end{itemize}
We do not impose the noise power constraint when retrieving the maximum privacy under perfect utility. Each data point shown in the figures is averaged over 100 independent experiments.
\subsection{Maximum privacy under perfect utility}

\begin{figure}[!htb]
	\centering
	\includegraphics[width=0.4\textwidth]{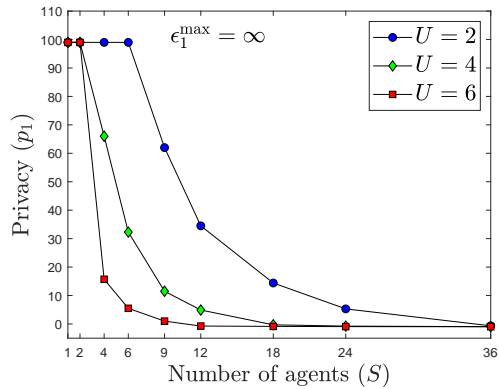}
	\caption{The maximum privacy under perfect utility when no prior information is available. Because the maximum privacy can be unbounded, it is truncated when exceeding 100.}
	\label{fig:maximum_privacy_no_prior}
\end{figure}
\begin{figure}[!htb]
	\centering
	\includegraphics[width=0.4\textwidth]{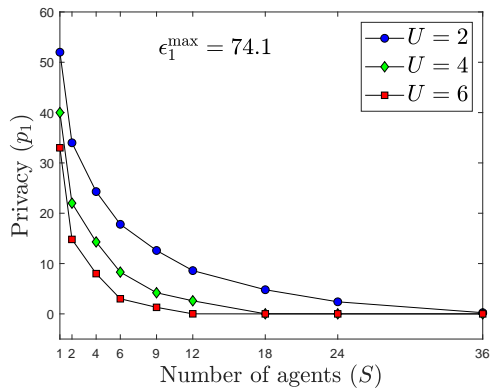}
	\caption{The maximum privacy under perfect utility when prior information is available. The maximum privacy threshold is $74.1$.}
	\label{fig:maximum_privacy_with_prior}
\end{figure}

We vary $S$ (the number of the agents) and $U$ (the dimension of the public parameter vector) to investigate the impact of these parameters on the maximum privacy attainable under perfect utility by solving \cref{eq:constraint_privacy}. Recall that $\epsilon_1^{\max} = \infty$ when $\bJ_0 = \bzero_L$ while $\epsilon_1^{\max}$ is bounded when $\bJ_0 \succ \bzero_L$. From \cref{fig:maximum_privacy_no_prior} with $\bJ_0 = \bzero_L$, it can be seen that the maximum privacy under perfect utility goes to infinity (ASUP is achieved) when $S = 1$. By examining the necessary and sufficient conditions for ASUP in \cref{thm:asup_no_prior}, it is observed that a system with a smaller number of agents $S$ is more likely to fulfill the ASUP conditions than a system with a larger $S$ with all the other settings remaining the same. In \cref{fig:maximum_privacy_with_prior} with $\bJ_0 \succ \bzero_L$, none of the maximum privacy obtained under perfect utility is close to $\epsilon_1^{\max}$ (ASUP is not achieved). This can be elucidated from \cref{thm:asup_with_prior}, where every agent is required to satisfy certain orthogonality conditions for ASUP to be achievable. Thus it is unlikely to generate a random system model to achieve ASUP. Furthermore, both \cref{fig:maximum_privacy_no_prior,fig:maximum_privacy_with_prior} demonstrate a descending trend for the maximum privacy achievable as the number of agents $S$ increases or the length of the public parameter vector $\bu$ increases. To expound further on this, recall from \cref{thm:asup_no_prior,thm:asup_with_prior} that perfect utility is achieved if and only if
\begin{align*}
\trace{\sum_{i=1}^{S} \bU\subs{\bPsi}{:,\calS_i} \bTheta_{i} \subs{\bPsi}{:,\calS_i}\T\bU\T} = 0.
\end{align*}
Therefore, to obtain $u \left(\bI_N, \bTheta \right) = 0$, a larger $S$ or $U$ is likely to force more agents to have $\bTheta_{i} = \bzero_{N_i}$. Thus a system with a larger $S$ or $U$ has less degrees-of-freedom for privacy perturbation when maintaining perfect utility compared to a system with a smaller $S$ or $U$.

\subsection{Alternating optimization}
\begin{figure}[!htb]
	\centering
	\includegraphics[width=0.4\textwidth]{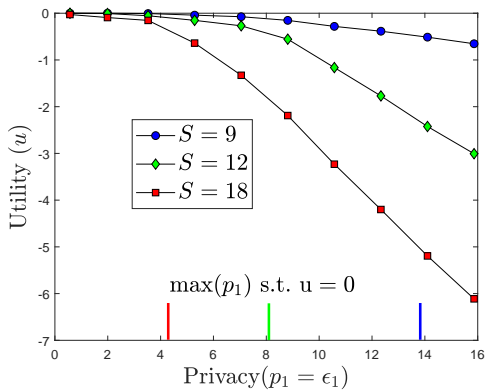}
	\caption{The privacy and utility tradeoff when $S=9,12,18$, respectively, obtained by using \cref{algo:sequential_opt}. The corresponding maximum privacy under perfect utility obtained by solving \cref{eq:constraint_privacy} is indicated.}
	\label{fig:utility_privacy_tradeoff}
\end{figure}

\begin{figure}[!htb]
	\centering
	\includegraphics[width=0.4\textwidth]{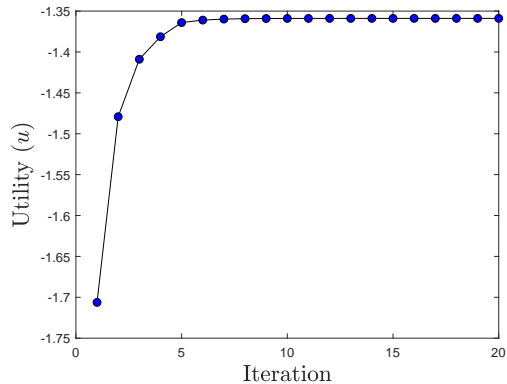}
	\caption{The utility at each iteration by using \cref{algo:sequential_opt}, where the privacy threshold is set to be 10.}
	\label{fig:sequential_iteration}
\end{figure}

By varying $S$ and the privacy threshold $\epsilon_1$, \cref{fig:utility_privacy_tradeoff} demonstrates the utility and privacy tradeoff by using \cref{algo:sequential_opt} proposed in \cref{sect:sequential_optimization}. It is observed that the utility decreases as the privacy increases. Under the same setting, a larger number of agents deteriorates the utility. This is because each agent contains less measurements, thus giving less degrees-of-freedom for the optimization and leading to smaller utility. Furthermore, it can be seen that the optimal utility obtained by the algorithm is close to perfect utility ($u=0$) when the privacy threshold $\epsilon_1$ is set to be equal to the maximum privacy obtained under perfect utility. This implies that the algorithm proposed in \cref{sect:sequential_optimization} approximates the optimal solution well. \cref{fig:sequential_iteration} shows the convergence of the algorithm.

%---------------------------------------------------------------------------%
%                            Sect:  Conclusion                              %
%---------------------------------------------------------------------------%
\section{Conclusion} \label{sect:conclusion}
We defined utility and privacy using the \gls{CRLB} and investigated the utility-privacy tradeoff in a decentralized setting. We showed that the privacy mechanism of linear compression can be arbitrarily closely approximated by the privacy mechanism using noise perturbation, and we showed how to translate between these two mechanisms. Furthermore, we derived necessary and sufficient conditions for achieving arbitrarily strong utility-privacy tradeoff, and developed methods to minimize privacy leakage while maintaining perfect utility under these conditions. When arbitrarily strong utility-privacy tradeoff is not achievable, we proposed an alternating optimization approach to find the optimal privacy noise power to add. Simulation results demonstrated the efficacy of our approach.

In this paper, we have considered a linear system model as well as linear privacy mechanisms. Analysis of nonlinear models and privacy mechanisms is much more challenging and is an interesting future research direction. In a nonlinear model, one possible method is to apply data-driven approaches like deep learning techniques to learn the privacy mechanisms similar to \cite{ChoPetXia:J2018} and \cite{ChoPetXia:J2017}, which however does not consider the scenario with explicit public and private parameters.

%---------------------------------------------------------------------------%
%                               Appendix                                    %
%---------------------------------------------------------------------------%
\appendices

\section{Derivation of \cref{eq:decomposed_eq_no_prior,eq:perturbed_CRLB_with_prior,eq:decomposed_eq_with_prior}} \label[Appendix]{appendix:D}
If $\bJ_0=\bzero_L$, we have
\begin{align} \nonumber
&\tbP_{\bx}(\bI_N, \bTheta) \\ \nonumber
&= \inv{\bH\T \inv{\bR + \bTheta} \bH} \\ \label{eq:decom_crlb_1}
&= \inv{\bH\T \left(\bR^{-1} - \bR^{-1} \inv{ \bI_N + \bTheta \bR^{-1} } \bTheta \bR^{-1} \right) \bH} \\ \nonumber
&= \inv{\bH\T \bR^{-1} \bH - \bH\T \bR^{-1} \inv{ \bI_N + \bTheta \bR^{-1} } \bTheta \bR^{-1} \bH} \\ 
\begin{split}\label{eq:decom_crlb_2}
&= \bP_\bx - \bP_\bx \bH\T \bR^{-1} \left(-\bI_N - \bTheta \bR^{-1} \right. \\
&\qquad\qquad\qquad\qquad \left. + \bTheta\bR^{-1} \bH \bP_\bx \bH\T \bR^{-1}\right)^{-1} \bTheta \bR^{-1} \bH \bP_\bx
\end{split} \\ \nonumber
&= \bP_\bx + \bPsi \inv{\bI_N + \bTheta \bPhi} \bTheta \bPsi\T,
\end{align}
where $\bPsi = \bP_\bx \bH\T \bR^{-1}$ and $\bPhi = \bR^{-1} - \bR^{-1} \bH \bP_\bx \bH\T \bR^{-1} \in \bbS^N$, \cref{eq:decom_crlb_1} follows from the binomial inverse theorem, and \cref{eq:decom_crlb_2} follows from the Woodbury matrix identity. Note that $\bPhi$ is a degenerate matrix since by letting the singular value decomposition of $\bH\T\bR^{-1/2} = \bU\bSigma\bV\T$, we have
\begin{align}
\bPhi
&= \bR^{-1} - \bR^{-1} \bH \bP_\bx \bH\T \bR^{-1} \nonumber \\
&= \bR^{-1} - \bR^{-1/2}\bR^{-1/2}\bH \nonumber \\ 
&\qquad\qquad\qquad \inv{\bH\T\bR^{-1/2}\bR^{-1/2}\bH}\bH\T\bR^{-1/2}\bR^{-1/2} \nonumber \\
&= \bR^{-1/2} \bV \diag(\bzero_L, \bI_{N-L}) \bV\T  \bR^{-1/2}. \label{eqn:J0_Phi}
\end{align}

On the other hand, if $\bJ_0\succ\bzero_L$, we obtain
\begin{align} \nonumber
&\tbP_{\bx}\left( \bI_N, \bTheta\right) \\ \nonumber
&= \inv{\bJ_0 + \bH\T \inv{\bR + \bTheta} \bH } \\ \label{eq:decom_crlb_4}
&= \bP_0 - \bP_0 \bH\T \inv{\bPhi^{-1} + \bTheta} \bH \bP_0 \\ \label{eq:decom_crlb_5}
&= \bP_0 - \bP_0 \bH\T \left( \bPhi - \bPhi \inv{\bI_N + \bTheta \bPhi}\bTheta \bPhi \right) \bH \bP_0 \\  \nonumber
&= \bP_0 - \bP_0 \bH\T \bPhi \bH\T \bP_0 + \bP_0 \bH\T \bPhi \inv{\bI_N + \bTheta \bPhi}\bTheta \bPhi\bH\bP_0 \\ \nonumber
&= \inv{\bJ_0 + \bH\T \bR^{-1} \bH} + \bP_0 \bH\T \bPhi \inv{\bI_N + \bTheta \bPhi}\bTheta \bPhi\bH\bP_0 \\ \nonumber
&= \bP_\bx + \bPsi \inv{\bI_N + \bTheta \bPhi}\bTheta \bPsi\T, 
\end{align}
where $\bP_0 = \bJ_0^{-1}$, $\bPhi = \inv{\bH\bP_0\bH\T + \bR}$, $\bPsi = \bP_0 \bH\T \bPhi$, \cref{eq:decom_crlb_4} follows from the Woodbury matrix identity, and \cref{eq:decom_crlb_5} is a consequence of the binomial inverse theorem.
\section{Proof of \cref{lemma:trace_matrix_inverse_equivalence}} \label[Appendix]{sec:lemma_eqiv}

Since $\trace{\bA \bS_n^{-1} \bA\T} \geq \trace{\bA \inv{\bS_n + \bB} \bA\T} \geq 0$, we have
\begin{align*}
&\lim_{n \to \infty}\trace{\bA \bS_n^{-1} \bA\T} = 0 \\
&\implies 
\lim_{n \to \infty}\trace{\bA \inv{\bS_n + \bB} \bA\T} = 0.
\end{align*}
On the other hand, for $\mu$ a positive scalar greater than the largest eigenvalue of $\bB$, we have
\begin{align*}
\trace{\bA \inv{\bS_n + \bB} \bA\T} 
\geq 
\trace{\bA \inv{\bS_n + \mu\bI_N} \bA\T} 
\geq 0.
\end{align*}
Therefore, 
\begin{align} \label{eq:lemma_eqiv_eq1}
\begin{split}
&\lim_{n \to \infty} \trace{\bA \inv{\bS_n + \bB} \bA\T} = 0 \\
&\implies 
\lim_{n \to \infty} \trace{\bA \inv{\bS_n + \mu\bI} \bA\T} = 0.
\end{split}
\end{align}
Let $\lambda_i(n) > 0$ be the $i$-th eigenvalue of $\bS_n$, associated with a unit eigenvector $\bv_i$ (commuting matrices share the same eigenspaces).
Let $\alpha_i = \norm{\bA\bv_i}_2^2$. We obtain
\begin{align*}
0 = \lim_{n \to \infty} \trace{\bA \inv{\bS_n + \mu\bI_N} \bA\T} 
= \lim_{n \to \infty} \sum_{i=1}^{N} \frac{\alpha_i}{\lambda_i(n) + \mu},
\end{align*}
hence either $\alpha_i = 0$ or $\lambda_i(n) \to \infty$ as $n \to \infty$ for each $i=1,\ldots,N$, which implies
\begin{align*}
\lim_{n \to \infty} \trace{\bA \bS_n^{-1} \bA\T} 
= \lim_{n \to \infty} \sum_{i=1}^{N} \frac{\alpha_i}{\lambda_i(n)} = 0.
\end{align*}
Combining the above result with \cref{eq:lemma_eqiv_eq1}, we obtain
\begin{align*}
&\lim_{n \to \infty} \trace{\bA \inv{\bS_n + \bB} \bA\T} = 0 \\
&\implies 
\lim_{n \to \infty} \trace{\bA \bS_n^{-1} \bA\T} = 0.
\end{align*}
The proof is now complete.

\section{Proof of \cref{lemma:vector_orthogonality}} \label[Appendix]{sec:vector_orthogonality}
Let $\lambda_i \geq 0$ be the $i$-th eigenvalue of $\bS$, associated with a unit eigenvector $\bv_i$. We first note that
\begin{align}
&\trace{\bA \bS \bA\T}
= \sum_{i=1}^N \lambda_i \norm{\bA\bv_i}_2^2, \label{ASA}\\
\begin{split}
&\lim_{n \to \infty} \trace{\bB \inv{n\bS + \frac{1}{n}\bI_N} \bB\T} \\
&= \lim_{n \to \infty}\sum_{i=1}^N \frac{1}{n\lambda_i - 1/n} \norm{\bB\bv_i}_2^2. \label{BSB}
\end{split}
\end{align}

We first prove necessity. If both \cref{ASA} and \cref{BSB} are $0$, $\bA \bv_i = \bzero$ if $\lambda_i\ne 0$, and $\bB \bv_i = \bzero$ if $\lambda_i = 0$ for each $i=1, \ldots, N$. This implies that $\bv_i$ is in the null space of either $\bA$, $\bB$ or both. Therefore, we have $\bA \bQ \bQ\T \bB\T = \bzero$, where $\bQ = \left[\bv_1, \ldots, \bv_N \right]$ is a unitary matrix. Thus, $\bA\bB\T = \bzero$.

We next assume that $\bA\bB\T = \bzero$. Let $\set{\bv_i}|{i=1,\ldots,\abs{\bB}}$ be a basis of the row space of $\bB$ and $\set{\bv_i}|{i= \abs{\bB}+1,\ldots, \abs{\bB}+\abs{\bA}}$ be a basis of the row space of $\bA$, where $\abs{\bA}=\rank(\bA)$ and $\abs{\bB}=\rank(\bB)$. Let $\lambda_i > 0$ if $i = 1, \ldots, \abs{\bB}$, else $\lambda_i = 0$. Let $\bS = \bQ\diag\parens*{(\lambda_i)_{i=1}^N}\bQ\T$, where $\bQ = \left[\bv_1, \ldots, \bv_N \right]$ is a unitary matrix. It can be easily verified that such $\bS$ satisfies the lemma conditions, and the proof is complete.
\section{Derivation of \cref{alt_u,alt_p}} \label[Appendix]{appendix:E}
We show the steps to obtain the utility and privacy function \gls{wrt} $\overline{\bTheta}_i$ described in \cref{sect:sequential_optimization}. Partition $\bPhi_i$ as
\begin{align*}
\bPhi_i = \begin{bmatrix} \subs{\bPhi_i}{\calN_i,\calN_i} & \subs{\bPhi_i}{\calN_i,\calN_{\bslash i}} \\ 
				\subs{\bPhi_i}{\calN_{\bslash i},\calN_i} & \subs{\bPhi_i}{\calN_{\bslash i},\calN_{\bslash i}} \end{bmatrix},
\end{align*}
where $\subs{\bPhi_i}{\calN_i,\calN_i} \in \bbR^{N_i \times N_i}$, $\calN_i = \{1, \ldots, N_i\}$, $\calN_{\bslash i} = \{N_i+1, \ldots, N\}$.
We have
\begin{align*}
&\inv{\diag(\overline{\bTheta}_{i}, \overline{\bTheta}_{\bslash{i}}) + \bPhi_i} \\
&= \begin{bmatrix} \overline{\bTheta}_{i} + \subs{\bPhi_i}{\calN_i,\calN_i} & \subs{\bPhi_i}{\calN_i,\calN_{\bslash i}} \\ 
\subs{\bPhi_i}{\calN_{\bslash i},\calN_i} & \overline{\bTheta}_{\bslash{i}} + \subs{\bPhi_i}{\calN_{\bslash i},\calN_{\bslash i}} \end{bmatrix}^{-1} \\
&= \begin{bmatrix} \bA_{11} & \bA_{12} \\ \bA_{21} & \bA_{22} \end{bmatrix},
\end{align*}
where 
\begin{align*}
\bA_{11} &= \inv{\overline{\bTheta}_{i} + \bOmega }, \\
\bA_{12} &= -\bA_{11} \subs{\bPhi_i}{\calN_i,\calN_{\bslash i}} \bT, \\
\bA_{21} &= -\bT \subs{\bPhi_i}{\calN_{\bslash i},\calN_i} \bA_{11}, \\
\bA_{22} &= \bT + \subs{\bPhi_i}{\calN_{\bslash i},\calN_i} \bA_{11} \subs{\bPhi_i}{\calN_i,\calN_{\bslash i}} \bT,
\end{align*}
where $\bOmega = \subs{\bPhi_i}{\calN_i,\calN_i} - \subs{\bPhi_i}{\calN_i,\calN_{\bslash i}} \bT \subs{\bPhi_i}{\calN_{\bslash i},\calN_i}$ and $\bT = \inv{ \overline{\bTheta}_{\bslash{i}} + \subs{\bPhi_i}{\calN_{\bslash i},\calN_{\bslash i}} }$. Therefore, we have
\begin{align*} 
&\overline{\bP}_{\bx} \left(\overline{\bTheta}_{i}\right) \\
&= \bP_\bx + \left[\subs{\bPsi_i}{:,\calN_i}, \subs{\bPsi_i}{:,\calN_{\bslash{i}}}\right] \\
&\qquad\qquad\qquad \begin{bmatrix} \bA_{11} & \bA_{12} \\ \bA_{21} & \bA_{22} \end{bmatrix} 
	\left[\subs{\bPsi_i}{:,\calN_i}, \subs{\bPsi_i}{:,\calN_{\bslash{i}}}\right]\T \\
&= \bP_\bx + \subs{\bPsi_i}{:,\calN_1} \bA_{11} \subs{\bPsi_i}{:,\calN_i}\T + 
	\subs{\bPsi_i}{:,\calN_{\bslash{i}}} \bA_{21} \subs{\bPsi_i}{:,\calN_i}\T  \\ 
	& \quad \quad \quad + \subs{\bPsi_i}{:,\calN_1} \bA_{12} \subs{\bPsi_i}{:,\calN_{\bslash{i}}}\T 
	+ \subs{\bPsi_i}{:,\calN_{\bslash{i}}} \bA_{22} \subs{\bPsi_i}{:,\calN_{\bslash{i}}}\T.
\end{align*}
Therefore, the utility and $j$-th privacy function can be written as
\begin{align*}
&u \left(\overline{\bTheta}_{i}\right)
= - \trace{\bGamma_u \inv{\overline{\bTheta}_{i} + \bOmega}} / \trace{\bU\bP_\bx\bU\T} - \bDelta_u,\\
&{p_j} \left(\overline{\bTheta}_{i}\right)
=  \trace{\bGamma_{p_j} \inv{\overline{\bTheta}_{i} + \bOmega}} / \trace{\bG_j\bP_\bx\bG_j\T} + \bDelta_{p_j},
\end{align*}
where $\bDelta_{p_j}, \bDelta_u, \bGamma_u, \bGamma_p$ are defined in \cref{sect:sequential_optimization}.

%---------------------------------------------------------------------------%
%                               Reference                                   %
%---------------------------------------------------------------------------%
\bibliographystyle{IEEEtran}
\bibliography{IEEEabrv,StringDefinitions,BibBooks,refs}

\end{document}

%% file: preamble.tex
\usepackage[T1]{fontenc}
\usepackage{amsmath,amssymb,amsfonts,mathrsfs,bm}% Typical maths resource packages
\usepackage{mathtools}
\usepackage{amsthm}
\usepackage{nicefrac}
\usepackage{cite}
\usepackage[shortlabels]{enumitem}
\usepackage{graphicx}
\usepackage{epstopdf}
\usepackage{url}
\usepackage{colortbl}
\usepackage{booktabs}
\usepackage{multirow}
\usepackage[table,dvipsnames]{xcolor}
\usepackage[normalem]{ulem}

\usepackage{array}
\newcolumntype{L}[1]{>{\raggedright\let\newline\\\arraybackslash\hspace{0pt}}m{#1}}
\newcolumntype{C}[1]{>{\centering\let\newline\\\arraybackslash\hspace{0pt}}m{#1}}
\newcolumntype{R}[1]{>{\raggedleft\let\newline\\\arraybackslash\hspace{0pt}}m{#1}}

\makeatletter
\let\MYcaption\@makecaption
\makeatother
\usepackage[font=footnotesize]{subcaption}
\makeatletter
\let\@makecaption\MYcaption
\makeatother

\usepackage{xparse}



\usepackage{glossaries}

\makeatletter
\let\oldgls\gls
\let\oldglspl\glspl

\newcommand\fussy@ifnextchar[3]{%
  \let\reserved@d=#1%
  \def\reserved@a{#2}%
  \def\reserved@b{#3}%
  \futurelet\@let@token\fussy@ifnch}
\def\fussy@ifnch{%
  \ifx\@let@token\reserved@d
    \let\reserved@c\reserved@a 
  \else
    \let\reserved@c\reserved@b
  \fi
 \reserved@c}

\renewcommand{\gls}[1]{%
  \oldgls{#1}\fussy@ifnextchar.{\@checkperiod}{\@}}
\renewcommand{\glspl}[1]{%
  \oldglspl{#1}\fussy@ifnextchar.{\@checkperiod}{\@}}

\newcommand{\@checkperiod}[1]{%
  \ifnum\sfcode`\.=\spacefactor\else#1\fi
}
\makeatother

\newacronym{wrt}{w.r.t.}{with respect to}
\newacronym{RHS}{RHS}{right-hand side}
\newacronym{LHS}{LHS}{left-hand side}
\newacronym{iid}{i.i.d.}{independent and identically distributed}
\usepackage{float}

\ifx\notloadhyperref\undefined
	\ifx\loadbibentry\undefined
		\usepackage[hidelinks,hypertexnames=false]{hyperref} 
	\else
		\usepackage{bibentry}
		\makeatletter\let\saved@bibitem\@bibitem\makeatother
		\usepackage[hidelinks,hypertexnames=false]{hyperref}
		\makeatletter\let\@bibitem\saved@bibitem\makeatother
	\fi
\else
	\relax
\fi

\usepackage[capitalize]{cleveref}
\crefname{equation}{}{}
\Crefname{equation}{}{}
\crefname{claim}{claim}{claims}
\crefname{step}{step}{steps}
\crefname{line}{line}{lines}
\crefname{condition}{condition}{conditions}
\crefname{dmath}{}{}
\crefname{dseries}{}{}
\crefname{dgroup}{}{}

\crefname{Problem}{Problem}{Problems}
\crefformat{Problem}{Problem~(#2#1#3)}
\crefrangeformat{Problem}{Problems~(#3#1#4) to~(#5#2#6)}

\crefname{Theorem}{Theorem}{Theorems}
\crefname{Corollary}{Corollary}{Corollaries}
\crefname{Proposition}{Proposition}{Propositions}
\crefname{Lemma}{Lemma}{Lemmas}
\crefname{Definition}{Definition}{Definitions}
\crefname{Example}{Example}{Examples}
\crefname{Assumption}{Assumption}{Assumptions}
\crefname{Remark}{Remark}{Remarks}
\crefname{Rem}{Remark}{Remarks}
\crefname{remarks}{Remarks}{Remarks}
\crefname{Appendix}{Appendix}{Appendices}
\crefname{Exercise}{Exercise}{Exercises}
\crefname{Theorem_A}{Theorem}{Theorems}
\crefname{Corollary_A}{Corollary}{Corollaries}
\crefname{Proposition_A}{Proposition}{Propositions}
\crefname{Lemma_A}{Lemma}{Lemmas}
\crefname{Definition_A}{Definition}{Definitions}

\usepackage{algorithm,algorithmic}

\ifx\loadbreqn\undefined
	\relax
\else
	\usepackage{breqn} 
\fi

\ifx\renewtheorem\undefined
\ifx\useTheoremCounter\undefined
\newtheorem{Theorem}{Theorem}
\newtheorem{Corollary}{Corollary}
\newtheorem{Proposition}{Proposition}
\newtheorem{Lemma}{Lemma}
\else
\newtheorem{Theorem}{Theorem}

\newtheorem{Proposition}[theorem]{Proposition}
\fi

\newtheorem{Definition}{Definition}
\newtheorem{Example}{Example}

\fi

\theoremstyle{remark}

\theoremstyle{plain}

\newcommand{\Real}{\mathbb{R}}

\newcommand{\iu}{\mathfrak{i}\mkern1mu}

\newcommand{\calA}{\mathcal{A}}
\newcommand{\calB}{\mathcal{B}}
\newcommand{\calC}{\mathcal{C}}

\newcommand{\calL}{\mathcal{L}}

\newcommand{\calN}{\mathcal{N}}

\newcommand{\calP}{\mathcal{P}}
\newcommand{\calQ}{\mathcal{Q}}

\newcommand{\calS}{\mathcal{S}}
\newcommand{\calT}{\mathcal{T}}

\newcommand{\bA}{\mathbf{A}}

\newcommand{\bB}{\mathbf{B}}

\newcommand{\bC}{\mathbf{C}}

\newcommand{\bg}{\mathbf{g}}
\newcommand{\bG}{\mathbf{G}}

\newcommand{\bH}{\mathbf{H}}

\newcommand{\bI}{\mathbf{I}}

\newcommand{\bJ}{\mathbf{J}}

\newcommand{\bK}{\mathbf{K}}

\newcommand{\bn}{\mathbf{n}}

\newcommand{\bP}{\mathbf{P}}

\newcommand{\bQ}{\mathbf{Q}}

\newcommand{\bR}{\mathbf{R}}

\newcommand{\bS}{\mathbf{S}}

\newcommand{\bT}{\mathbf{T}}
\newcommand{\bu}{\mathbf{u}}
\newcommand{\bU}{\mathbf{U}}
\newcommand{\bv}{\mathbf{v}}
\newcommand{\bV}{\mathbf{V}}
\newcommand{\bw}{\mathbf{w}}
\newcommand{\bW}{\mathbf{W}}
\newcommand{\bx}{\mathbf{x}}
\newcommand{\bX}{\mathbf{X}}
\newcommand{\by}{\mathbf{y}}
\newcommand{\bY}{\mathbf{Y}}

\newcommand{\bZ}{\mathbf{Z}}

\newcommand{\bbR}{\mathbb{R}}
\newcommand{\bbS}{\mathbb{S}}

\DeclareSymbolFont{bsfletters}{OT1}{cmss}{bx}{n}
\DeclareSymbolFont{ssfletters}{OT1}{cmss}{m}{n}
\DeclareMathSymbol{\bsfGamma}{0}{bsfletters}{'000}
\DeclareMathSymbol{\ssfGamma}{0}{ssfletters}{'000}
\DeclareMathSymbol{\bsfDelta}{0}{bsfletters}{'001}
\DeclareMathSymbol{\ssfDelta}{0}{ssfletters}{'001}
\DeclareMathSymbol{\bsfTheta}{0}{bsfletters}{'002}
\DeclareMathSymbol{\ssfTheta}{0}{ssfletters}{'002}
\DeclareMathSymbol{\bsfLambda}{0}{bsfletters}{'003}
\DeclareMathSymbol{\ssfLambda}{0}{ssfletters}{'003}
\DeclareMathSymbol{\bsfXi}{0}{bsfletters}{'004}
\DeclareMathSymbol{\ssfXi}{0}{ssfletters}{'004}
\DeclareMathSymbol{\bsfPi}{0}{bsfletters}{'005}
\DeclareMathSymbol{\ssfPi}{0}{ssfletters}{'005}
\DeclareMathSymbol{\bsfSigma}{0}{bsfletters}{'006}
\DeclareMathSymbol{\ssfSigma}{0}{ssfletters}{'006}
\DeclareMathSymbol{\bsfUpsilon}{0}{bsfletters}{'007}
\DeclareMathSymbol{\ssfUpsilon}{0}{ssfletters}{'007}
\DeclareMathSymbol{\bsfPhi}{0}{bsfletters}{'010}
\DeclareMathSymbol{\ssfPhi}{0}{ssfletters}{'010}
\DeclareMathSymbol{\bsfPsi}{0}{bsfletters}{'011}
\DeclareMathSymbol{\ssfPsi}{0}{ssfletters}{'011}
\DeclareMathSymbol{\bsfOmega}{0}{bsfletters}{'012}
\DeclareMathSymbol{\ssfOmega}{0}{ssfletters}{'012}

\newcommand{\bxi}{\bm{\xi}}

\newcommand{\bGamma}{\bm{\Gamma}}
\newcommand{\bLambda}{\bm{\Lambda}}
\newcommand{\bSigma	}{\bm{\Sigma}}
\newcommand{\bPsi}{\bm{\Psi}}
\newcommand{\bDelta}{\bm{\Delta}}
\newcommand{\bXi}{\bm{\Xi}}

\newcommand{\bOmega}{\bm{\Omega}}
\newcommand{\bPhi}{\bm{\Phi}}

\newcommand{\bTheta}{\bm{\Theta}}

\newcommand{\tbTheta}{\widetilde{\bTheta}}

\DeclareMathOperator{\st}{s.t.}

\DeclareMathOperator{\diag}{diag}

\DeclareMathOperator{\Tr}{Tr}

\DeclareMathOperator{\cov}{cov}

\DeclareMathOperator{\rank}{rank}

\DeclarePairedDelimiter\abs{\lvert}{\rvert}
\DeclarePairedDelimiter\parens{(}{)}

\newcommand{\qednew}{\nobreak \ifvmode \relax \else
      \ifdim\lastskip<1.5em \hskip-\lastskip
      \hskip1.5em plus0em minus0.5em \fi \nobreak
      \vrule height0.75em width0.5em depth0.25em\fi}

\newcommand{\nn}{\nonumber\\}

\newcommand{\T}{^{\intercal}}% transpose notation
\newcommand{\bzero}{\mathbf{0}}

\newcommand{\norm}[1]{{\left\lVert{#1}\right\rVert}}
\newcommand{\trace}[1]{{\Tr\left( #1 \right)}}
\DeclareDocumentCommand\set{s m t| m}{%
  \IfBooleanTF#1%
	{\left\{\, #2\mathrel{} \IfBooleanTF{#3}{\middle|}{:}\mathrel{}  #4\, \right\}}%
  {\{\, #2 \IfBooleanTF{#3}{\mid}{\mathrel{} : \mathrel{}} #4\, \}}% 
}

\DeclareDocumentCommand \ifcond {m m} {%
	{#1} %
	\IfValueT{#2}{\, \middle|\, {#2}}%
}

\DeclareDocumentCommand \P {e{_} g >{\SplitArgument{ 1 }{ @| }}d() g } {%
	\mathbb{P}%
	\IfValueTF{#1}{_{#1}}
		{\IfValueT{#2}{_{#2}}}%
	\IfValueT{#3}{\left(\ifcond#3}%
	\IfValueT{#4}{\, \middle|\, {#4}}%
	\IfValueT{#3}{\right)}%
}

\DeclareDocumentCommand \E {e{_} g >{\SplitArgument{ 1 }{ @| }}o g } {%
	\mathbb{E}%
	\IfValueTF{#1}{_{#1}}
		{\IfValueT{#2}{_{#2}}}%
	\IfValueT{#3}{\left[\ifcond#3}%
	\IfValueT{#4}{\, \middle|\, {#4}}%
	\IfValueT{#3}{\right]}%
}

\newcommand{\Unif}[1]{\mathrm{Unif}\left(#1\right)}

\newcommand{\N}[2]{{\calN\left({#1},\, {#2}\right)}}

\definecolor{gray90}{gray}{0.9}

\ifx\nohighlights\undefined

	\newcommand{\msout}[1]{\text{\color{green} \sout{\ensuremath{#1}}}}
	\newcommand{\del}[1]{{\color{green}\ifmmode \msout{#1}\else\sout{#1}\fi}}
\else

	\newcommand{\msout}[1]{#1}
	\newcommand{\del}[1]{#1}
\fi

\newcommand{\hhide}[1]{}
\graphicspath{{./Figures/},{./BioPhotos/}} 
\ifx\diagnoselabel\undefined
	\relax
\else
	\makeatletter
	 \def\@testdef #1#2#3{%
		 \def\reserved@a{#3}\expandafter \ifx \csname #1@#2\endcsname
		\reserved@a  \else
	 \typeout{^^Jlabel #2 changed:^^J%
	 \meaning\reserved@a^^J%
	 \expandafter\meaning\csname #1@#2\endcsname^^J}%
	 \@tempswatrue \fi}
	\makeatother
\fi